\documentclass[12pt]{article}
\usepackage[dvips]{graphicx}
\usepackage{epsfig}
\usepackage{latexsym,amsmath,amsfonts,amssymb}
\usepackage[latin1]{inputenc}
\usepackage[american]{babel}
\usepackage[dvips]{graphicx}
\usepackage{bbm}
\usepackage{color}
\usepackage[unicode]{hyperref}
\def\im{Invent. Math.}

\def\a{\alpha}
\def\b{\beta}
\def\c{\gamma}
\def\d{\delta}
\def\f{\phi}               
\def\vf{\varphi}  \def\tvf{\tilde{\varphi}}
\def\vp{\varphi}
\def\g{\gamma}
\def\h{\eta}
\def\j{\psi}
\def\k{\kappa}                    
\def\l{\lambda}
\def\m{\mu}
\def\n{\nu}
\def\o{\omega}  \def\w{\omega}
\def\p{\pi}
\def\q{\theta}  \def\th{\theta}                  
\def\r{\rho}                                     
\def\s{\sigma}                                   
\def\t{\tau}
\def\u{\upsilon}
\def\x{\xi}
\def\z{\zeta}
\def\pt{\tilde{\varphi}}
\def\tt{\tilde{\theta}}
\def\lab{\label}
\def\6{\partial}
\def\wg{\wedge}
\def\bpsi{\bar{\psi}}
\def\bt{\bar{\theta}}
\def\bvf{\bar{\varphi}}

\DeclareMathOperator{\tr}{tr}

\newcommand{\be}{\begin{equation}}
\newcommand{\ee}{\end{equation}}
\newcommand{\beq}{\begin{equation}}
\newcommand{\eeq}{\end{equation}}
\newcommand{\bea}{\begin{eqnarray}}
\newcommand{\eea}{\end{eqnarray}}
\newcommand{\nn}{\nonumber}

\newcommand{\ba}{\begin{eqnarray}}
\newcommand{\ea}{\end{eqnarray}}

\newcommand{\beqs}{\begin{eqnarray}}
\newcommand{\eeqs}{\end{eqnarray}}
\newcommand{\bal}{\begin{aligned}}
\newcommand{\eal}{\end{aligned}}

\begin{document}
\baselineskip=15.5pt
\pagestyle{plain}
\setcounter{page}{1}


\def\del{{\partial}}
\def\vev#1{\left\langle #1 \right\rangle}
\def\cn{{\cal N}}
\def\co{{\cal O}}
\def\IC{{\mathbb C}}
\def\IR{{\mathbb R}}
\def\IZ{{\mathbb Z}}
\def\RP{{\bf RP}}
\def\CP{{\bf CP}}
\def\Poincare{{Poincar\'e }}
\def\tr{{\rm tr}}
\def\tp{{\tilde \Phi}}

\def\TL{\hfil$\displaystyle{##}$}
\def\TR{$\displaystyle{{}##}$\hfil}
\def\TC{\hfil$\displaystyle{##}$\hfil}
\def\TT{\hbox{##}}
\def\HLINE{\noalign{\vskip1\jot}\hline\noalign{\vskip1\jot}}
\def\seqalign#1#2{\vcenter{\openup1\jot
   \halign{\strut #1\cr #2 \cr}}}
\def\lbldef#1#2{\expandafter\gdef\csname #1\endcsname {#2}}
\def\eqn#1#2{\lbldef{#1}{(\ref{#1})}%
\begin{equation} #2 \label{#1} \end{equation}}
\def\eqalign#1{\vcenter{\openup1\jot
     \halign{\strut\span\TL & \span\TR\cr #1 \cr
    }}}
\def\eno#1{(\ref{#1})}
\def\href#1#2{#2}
\def\half{\frac{1}{2}}

\def\ads{{\it AdS}}
\def\adsp{{\it AdS}$_{p+2}$}
\def\cft{{\it CFT}}

\newcommand{\ber}{\begin{eqnarray}}
\newcommand{\eer}{\end{eqnarray}}

\newcommand{\beqar}{\begin{eqnarray}}
\newcommand{\cN}{{\cal N}}
\newcommand{\cO}{{\cal O}}
\newcommand{\cA}{{\cal A}}
\newcommand{\cT}{{\cal T}}
\newcommand{\cF}{{\cal F}}
\newcommand{\cC}{{\cal C}}
\newcommand{\cR}{{\cal R}}
\newcommand{\cW}{{\cal W}}
\newcommand{\eeqar}{\end{eqnarray}}
\newcommand{\tht}{\thteta}
\newcommand{\lm}{\lambda}\newcommand{\Lm}{\Lambda}


\newcommand{\nonu}{\nonumber}
\newcommand{\oh}{\displaystyle{\frac{1}{2}}}
\newcommand{\dsl}
   {\kern.06em\hbox{\raise.15ex\hbox{$/$}\kern-.56em\hbox{$\partial$}}}
\newcommand{\id}{i\!\!\not\!\partial}
\newcommand{\as}{\not\!\! A}
\newcommand{\ps}{\not\! p}
\newcommand{\ks}{\not\! k}
\newcommand{\D}{{\cal{D}}}
\newcommand{\dv}{d^2x}
\newcommand{\Z}{{\cal Z}}
\newcommand{\N}{{\cal N}}
\newcommand{\Dsl}{\not\!\! D}
\newcommand{\Bsl}{\not\!\! B}
\newcommand{\Psl}{\not\!\! P}
\newcommand{\eeqarr}{\end{eqnarray}}
\newcommand{\ZZ}{{\rm \kern 0.275em Z \kern -0.92em Z}\;}


\def\del{{\delta^{\hbox{\sevenrm B}}}} \def\ex{{\hbox{\rm e}}}
\def\azb{A_{\bar z}} \def\az{A_z} \def\bzb{B_{\bar z}} \def\bz{B_z}
\def\czb{C_{\bar z}} \def\cz{C_z} \def\dzb{D_{\bar z}} \def\dz{D_z}
\def\im{{\hbox{\rm Im}}} \def\mod{{\hbox{\rm mod}}} \def\tr{{\hbox{\rm Tr}}}
\def\ch{{\hbox{\rm ch}}} \def\imp{{\hbox{\sevenrm Im}}}
\def\trp{{\hbox{\sevenrm Tr}}} \def\vol{{\hbox{\rm Vol}}}
\def\rl{\Lambda_{\hbox{\sevenrm R}}} \def\wl{\Lambda_{\hbox{\sevenrm W}}}
\def\fc{{\cal F}_{k+\cox}} \def\vev{vacuum expectation value}
\def\nodiv{\mid{\hbox{\hskip-7.8pt/}}}
\def\ie{{\em i.e.}}
\def\ie{\hbox{\it i.e.}}

\def\CC{{\mathchoice
{\rm C\mkern-8mu\vrule height1.45ex depth-.05ex
width.05em\mkern9mu\kern-.05em}
{\rm C\mkern-8mu\vrule height1.45ex depth-.05ex
width.05em\mkern9mu\kern-.05em}
{\rm C\mkern-8mu\vrule height1ex depth-.07ex
width.035em\mkern9mu\kern-.035em}
{\rm C\mkern-8mu\vrule height.65ex depth-.1ex
width.025em\mkern8mu\kern-.025em}}}

\def\RR{{\rm I\kern-1.6pt {\rm R}}}
\def\NN{{\rm I\!N}}
\def\ZZ{{\rm Z}\kern-3.8pt {\rm Z} \kern2pt}
\def\IB{\relax{\rm I\kern-.18em B}}
\def\ID{\relax{\rm I\kern-.18em D}}
\def\II{\relax{\rm I\kern-.18em I}}
\def\IP{\relax{\rm I\kern-.18em P}}
\newcommand{\CS}{{\scriptstyle {\rm CS}}}
\newcommand{\CSs}{{\scriptscriptstyle {\rm CS}}}
\newcommand{\rc}{\nonumber\\}
\newcommand{\bear}{\begin{eqnarray}}
\newcommand{\eear}{\end{eqnarray}}

\newcommand{\LL}{{\cal L}}

\def\mani{{\cal M}}
\def\calo{{\cal O}}
\def\calb{{\cal B}}
\def\calw{{\cal W}}
\def\calz{{\cal Z}}
\def\cald{{\cal D}}
\def\calc{{\cal C}}
\def\to{\rightarrow}
\def\ele{{\hbox{\sevenrm L}}}
\def\ere{{\hbox{\sevenrm R}}}
\def\zb{{\bar z}}
\def\wb{{\bar w}}
\def\nodiv{\mid{\hbox{\hskip-7.8pt/}}}
\def\menos{\hbox{\hskip-2.9pt}}
\def\dr{\dot R_}
\def\drr{\dot r_}
\def\ds{\dot s_}
\def\da{\dot A_}
\def\dga{\dot \gamma_}
\def\ga{\gamma_}
\def\dal{\dot\alpha_}
\def\al{\alpha_}
\def\cl{{closed}}
\def\cls{{closing}}
\def\vev{vacuum expectation value}
\def\tr{{\rm Tr}}
\def\to{\rightarrow}
\def\too{\longrightarrow}


\def\a{\alpha}
\def\b{\beta}
\def\c{\gamma}
\def\d{\delta}
\def\e{\epsilon}           
\def\F{\Phi}
\def\f{\phi}               
\def\vf{\varphi}  \def\tvf{\tilde{\varphi}}
\def\vp{\varphi}
\def\g{\gamma}
\def\h{\eta}
\def\j{\psi}
\def\k{\kappa}                    
\def\l{\lambda}
\def\m{\mu}
\def\n{\nu}
\def\o{\omega}  \def\w{\omega}
\def\q{\theta}  \def\th{\theta}                  
\def\r{\rho}                                     
\def\s{\sigma}                                   
\def\t{\tau}
\def\u{\upsilon}
\def\x{\xi}
\def\X{\Xi}
\def\z{\zeta}
\def\pt{\tilde{\varphi}}
\def\tt{\tilde{\theta}}
\def\lab{\label}
\def\6{\partial}
\def\wg{\wedge}
\def\atanh{{\rm arctanh}}
\def\bpsi{\bar{\psi}}
\def\bt{\bar{\theta}}
\def\bvf{\bar{\varphi}}

%

\newfont{\namefont}{cmr10}
\newfont{\addfont}{cmti7 scaled 1440}
\newfont{\boldmathfont}{cmbx10}
\newfont{\headfontb}{cmbx10 scaled 1728}
\newcommand{\re}{\,\mathbb{R}\mbox{e}\,}
\newcommand{\hyph}[1]{$#1$\nobreakdash-\hspace{0pt}}
\providecommand{\abs}[1]{\lvert#1\rvert}
\newcommand{\Nugual}[1]{$\mathcal{N}= #1 $}
\newcommand{\sub}[2]{#1_\text{#2}}
\newcommand{\partfrac}[2]{\frac{\partial #1}{\partial #2}}
\newcommand{\bsp}[1]{\begin{equation} \begin{split} #1 \end{split} \end{equation}}
\newcommand{\calF}{\mathcal{F}}
\newcommand{\calO}{\mathcal{O}}
\newcommand{\calM}{\mathcal{M}}
\newcommand{\calV}{\mathcal{V}}
\newcommand{\bbZ}{\mathbb{Z}}
\newcommand{\bbC}{\mathbb{C}}
\newcommand{\cK}{{\cal K}}

\newcommand{\Thq}{\Theta\left(\r-\r_q\right)}
\newcommand{\Dq}{\d\left(\r-\r_q\right)}
\newcommand{\kten}{\kappa^2_{\left(10\right)}}
\newcommand{\pbi}[1]{\imath^*\left(#1\right)}
\newcommand{\ho}{\hat{\omega}}
\newcommand{\tth}{\tilde{\th}}
\newcommand{\tf}{\tilde{\f}}
\newcommand{\tj}{\tilde{\j}}
\newcommand{\tw}{\tilde{\omega}}
\newcommand{\tz}{\tilde{z}}
\newcommand{\prj}[2]{(\partial_r{#1})(\partial_{\j}{#2})-(\partial_r{#2})(\partial_{\j}{#1})}
\def\atanh{{\rm arctanh}}
\def\sech{{\rm sech}}
\def\csch{{\rm csch}}
\allowdisplaybreaks[1]

\def\red{\textcolor[rgb]{0.98,0.00,0.00}}

\numberwithin{equation}{section}

\newcommand{\Tr}{\mbox{Tr}}    


%
\renewcommand{\theequation}{{\rm\thesection.\arabic{equation}}}
\begin{titlepage}

\hfill FPAUO-14/06  \\

\phantom{xx}
\vskip 0.4in

\begin{center}
{\Large \bf A New $AdS_4/CFT_3$ Dual with Extended SUSY and}
\vskip 0.1in

{\Large \bf  a Spectral Flow}

\vskip 0.4in

{\bf Yolanda Lozano}${}^{a,}{}^{1}$,~{\bf Niall T. Macpherson}${}^{a,b,}{}^{2}$
\\

\vskip .2in

a: Department of Physics,  University of Oviedo,\\
Avda.~Calvo Sotelo 18, 33007 Oviedo, Spain\\

\vskip .2in

b: Department of Physics, Swansea University\\
Singleton Park, Swansea SA2 8PP, United Kingdom\\
 
\vskip 5mm

\vspace{0.2in}
\end{center}
\vspace{0.2in}
\centerline{{\bf Abstract}}
We construct a new $AdS_4$ background in Type IIB supergravity by means of a non-Abelian T-duality transformation on the Type IIA dual of ABJM. The analysis of probe and particle-like branes  suggests a dual CFT in which each of the gauge groups is doubled. 
A common feature of non-Abelian T-duality is that in the absence of any global information coming from String Theory it gives rise to non-compact dual backgrounds, with coordinates living in the Lie algebra of  the Lie group involved in the dualization. In backgrounds with CFT duals this poses obvious problems to the CFTs.   In this paper we show that for the new $AdS_4$ background
the gauge groups of the associated dual CFT undergo a spectral flow as the non-compact internal direction runs from 0 to infinity, which resembles Seiberg duality in $\mathcal{N}=1$. This phenomenon,
very reminiscent of the cascade, provides an interpretation in the CFT for the running of the non-compact coordinate, and suggests that at the end of the flow the extra charges disappear and the dual CFT is described by a 2-node quiver very similar to ABJM, albeit with reduced supersymmetry.
\smallskip

\smallskip

\vfill
\noindent
 {
 $^1$ylozano@uniovi.es,
$^2$pymacpherson@swansea.ac.uk}

\end{titlepage}
\setcounter{footnote}{0}

\tableofcontents

\setcounter{footnote}{0}
\renewcommand{\theequation}{{\rm\thesection.\arabic{equation}}}

 \newpage

\section{Introduction}

Three dimensional Chern-Simons theories play an important role in condensed matter systems, where they are often strongly coupled. In view of this it is interesting to construct  new $AdS_4$ solutions with 3d CFT duals. Many CS-matter theories with IIA/M-theory gravity duals and varying supersymmetries have been studied \cite{Benna:2008zy}-\cite{Bergman:2010xd} 
starting with the $\mathcal{N}=6$ ABJM theory \cite{Aharony:2008ug} and its extension to unequal ranks \cite{Aharony:2008gk} \footnote{See also \cite{Gaiotto:2007qi} for previous work.}. In this paper we construct one further example in Type IIB 
through a non-Abelian T-duality transformation on the ABJM background, and study some of the properties of its dual CFT. This solution represents a generalization of the class of Type IIB solutions presented in \cite{Lust:2009mb} in which the internal manifold does not have a direct product structure.

Non-Abelian T-duality was first studied at the level of the string $\sigma$-model  in the early 90's, when the authors of  \cite{de la Ossa:1992vc}  extended to non-Abelian isometry groups the gauging procedure introduced in \cite{Rocek:1991ps} for deriving T-dual geometries for strings in curved backgrounds with Abelian isometries.
In spite of its very close analogy with Abelian T-duality, non-Abelian T-duality did not reach however the status of a string theory symmetry, for two main reasons. First, the derivation in \cite{de la Ossa:1992vc} only works for spherical worldsheets. Second, unlike its Abelian counterpart \cite{Rocek:1991ps}, non-Abelian T-duality is only known to respect conformal symmetry to first order in $\alpha^\prime$. 
From a more practical point of view we also lack a mechanism to extract global information about the dual background. 

Yet recent research in non-Abelian T-duality, which started with the derivation in  \cite{Sfetsos:2010uq} of the transformation rules of the RR potentials, has shown that it can be very useful in generating new supergravity backgrounds with CFT duals, even if corrections may be expected beyond the planar limit and strong coupling regimes. 
Interesting examples of non-Abelian T-duals (NATD) of $AdS$ spacetimes, such as a background closely related to the $\mathcal{N}=2$ Gaiotto-Maldacena geometries in
\cite{Gaiotto:2009gz}, derived through NATD from $AdS_5\times S^5$, were worked out in  \cite{Sfetsos:2010uq}. It was also shown  (see also \cite{Itsios:2012dc}) that NATD typically breaks supersymmetry by a half\footnote{When the isometry group acts without fixed points, while it is completely broken in the presence of fixed points \cite{Lozano:2011kb}.}, with the previous example providing a concrete realization of this.

The potential of non-Abelian T-duality as a solution generating technique was strengthened with the construction in \cite{Lozano:2012au} of a supersymmetry preserving $AdS_6$ solution to Type IIB, obtained  as the result of a non-Abelian T-duality transformation on the only known supersymmetric $AdS_6$ solution to massive Type IIA \cite{Brandhuber:1999np}. This example provided the second only $AdS_6$ solution known in Type IIB, where strong constraints imposed by supersymmetry rule out the existence of more ``standard'' solutions  \cite{Passias:2012vp},  besides the Abelian T-dual of the Brandhuber-Oz solution  \cite{Cvetic:2000cj,Lozano:2012au}.
It also provided the first NATD background with supersymmetry fully preserved. It was later shown in \cite{Lozano:2013oma} that this new solution can be of relevance to the study of new classes of 5d fixed point theories with gravity duals  \cite{Seiberg:1996bd,Intriligator:1997pq}. More recently, the conditions for the existence of $AdS_6\times M_4$
supersymmetry preserving solutions to Type IIB have been derived  \cite{Apruzzi:2014qva}, and even if no explicit solutions other than the just mentioned Abelian and non-Abelian T-duals of the Brandhuber-Oz background have been found, there are hints that other solutions such as those describing the $(p,q)$-five brane webs in \cite{Aharony:1997ju}  may also exist.

Other NATD backgrounds  with supersymmetry fully preserved and different applications in the context of the AdS/CFT correspondence have been derived in \cite{Itsios:2012zv}-\cite{Caceres:2014uoa}, with some of them exhibiting 
interesting properties of the original dual CFTs, 
such as Seiberg duality, domain walls, confinement or symmetry breaking.
In the context of 3d field theories, confining $\mathcal{N}=1$ Chern-Simons theories with multiple Chern-Simons levels have been generated in \cite{Macpherson:2013zba}.

In this paper we apply non-Abelian T-duality to the IIA background of ABJM and obtain a new $AdS_4$ solution in Type IIB with a CFT dual whose properties we study. The  paper is organized as follows. In section 2 we summarize the main features of the $AdS_4\times \mathbb{CP}^3$ background. In section 3 we present the NATD solution and discuss some of its most relevant properties. In order to describe the solution in terms of $b\in [0,1)$, with 
$b=\frac{1}{4\pi^2}|\int B_2\, | $, large gauge transformations are needed which define a range for a newly generated non-compact direction.  We show that two charges interpreted as ranks in the dual CFT as well as two charges interpreted as levels can be defined in the different ranges. 
We finally analyze the conditions for the Kosmann derivative to vanish. These, together with the $SO(2)$ isometries preserved by the background point at $\mathcal{N}=2$ supersymmetry. Section 4 is devoted to examining different properties of the 3d dual CFT, and contains the main results of the paper. We construct BPS color branes, responsible for the ranks of the dual gauge groups, as well as particle-like branes with tadpoles and/or conformal dimensions associated to the ranks and levels. The picture that arises is that the gauge groups of the original ABJM theory are doubled in the dual, with large gauge transformations inducing a flow in the charges that resembles Seiberg duality in $\mathcal{N}=1$. In section 4.3 we put
all the results together and make a proposal for the 3d dual CFT in which  the gauge groups undergo a spectral flow, very reminiscent of the cascade, as the non-compact internal direction runs from zero to infinity.  This flow is the reflection in the CFT
of the running of the non-compact direction, and suggests that  the field theories that are dual to the solution for its different ranges  could be equivalent to the field theory at the end of the flow. This theory seems to be a 2-node quiver very similar to ABJM, albeit with reduced $\mathcal{N}=2$ supersymmetry. In section 5 we construct various dual giant graviton configurations which should allow to prove the presence of mesonic operators built out of  bifundamental fields in the dual CFT. 
In section 6 we compute the central charge, entanglement entropy and free energy of the dual background and show that up to a constant, whose origin we explain, they reproduce the results in the original ABJM theory. Section 7 contains the Conclusions. Here we summarize the main results of the paper and speculate with a possible  IIB brane realization of the dual CFT. Finally, Appendix A collects the transformation rules of the RR fields under the NATD transformation and Appendix B contains the explicit derivation of the Killing spinors on our chosen parameterization of $\mathbb{CP}^3$.

\section{The IIA $AdS_4\times \mathbb{CP}^3$ solution}

We express the metric of $AdS_4\times \mathbb{CP}^3$ in string frame as
\begin{equation}\label{eq: metricbefore1} 
ds_{str}^2=\frac{L^2}{4} ds^2_{AdS_4}+L^2ds^2_{\mathbb{CP}^3}=
\frac{L^2}{4}\bigg( \rho^2 dx^2_{1,2}+\frac{d\rho^2}{\rho^2}\bigg)+L^2ds^2_{\mathbb{CP}^3}
\end{equation}
and parameterize the $ \mathbb{CP}^3$ in coordinates adapted to its foliation in $T^{1,1}=S^2\times S^3$ \cite{Cvetic:2000yp}
\beq\label{eq: metricbefore2} 
\begin{array}{l l}
ds^2_{\mathbb{CP}^3}&=d\zeta^2+\frac{1}{4}\bigg(\cos^2\zeta(d\theta_1^2+\sin^2\theta_1 d\phi_1^2)+\sin^2\zeta(d\theta_2^2+\sin^2\theta_2 d\phi_2^2)+\\
&~~~~~~~~~~~~~~~~~~~~~~~~~~~~~~~~~~~~+\sin^2\zeta\cos^2\zeta(d\psi+\cos\theta_1 d\phi_1+\cos\theta_2 d\phi_2)^2\bigg)\\
&=d\zeta^2+\frac{1}{4}\bigg(\cos^2\zeta(d\theta_1^2+\sin^2\theta_1 d\phi_1^2)+\sin^2{\zeta}(\o_1^2+\o_2^2)+\sin^2{\zeta}\cos^2{\zeta}(\o_3+\cos\theta_1 d\phi_1)^2\bigg)
\end{array}
\eeq
where $0\leq \zeta < \frac{\pi}{2},~0\leq \theta_i<\pi,~0\leq\phi_i\leq2\pi,~0\leq \psi \leq 4\pi$.

The background supports non-trivial 2 and 4-form RR fluxes
\beq
\begin{array}{l l}
F_2&=-\frac{k}{2} \bigg(\cos^2\zeta\sin\theta_1 d\theta_1\wedge d\phi_1-\sin^2\zeta\sin\theta_2 d\theta_2\wedge d\phi_2+\\
&~~~~~~~~~~~~~~~~~~~~~~~~~~+\sin 2\zeta d\zeta\wedge(d\psi+\cos\theta_1 d\phi_1+\cos\theta_2 d\phi_2 )\bigg),\\
F_4&= \frac38\,  k L^2 \rho^2 dt \wedge dx_1\wedge dx_2 \wedge d\rho \, .
\end{array}
\eeq
The dilaton is in turn given by
\beq
e^{\phi}=\frac{L}{k},
\eeq
and the radius of curvature $L$ by
\beq
\label{rank}
L^4=\frac{32  \pi^2 N}{k}\, .
\eeq
The dual quiver has two nodes of rank $N$, which are realised in the geometry via the D2 brane charge on $\mathbb{CP}^3$
\beq
-\frac{1}{32\pi^5}\int_{\mathbb{CP}^3}\star F_4 = N.
\eeq
In addition to this there are two Chern-Simons terms whose levels $k_1=-k_2=k$ are given by the D6 brane charge on $\mathbb{CP}^1$
\beq
\frac{1}{2\pi}\int_{\mathbb{CP}^1} F_2=k\, .
\eeq
As we will dualize on a freely acting $SU(2)$ isometry defined by $\psi$ and one of the two 2-spheres it would be useful to have a precise map that exchanges the spheres and does nothing else. That way one can perform this map in the dual and get the equivalent result. Such a map is given by
\beq
\zeta\to \zeta+\frac{\pi}{2},~~~~F_2\to- F_2,
\eeq
which one can check is completely equivalent to $(\theta_1,\phi_1)\leftrightarrow(\theta_2,\phi_2)$. 

\section{The IIB NATD $AdS_4$ solution}

The $AdS_4\times \mathbb{CP}^3$ background is invariant under freely acting $SU(2)$ isometries on 
$\psi$ and any of the two 2-spheres parameterized by $(\theta_1,\phi_1)$ and $(\theta_2,\phi_2)$, that we denote by $S^2_1$ and $S^2_2$.  We will perform the non-Abelian T-duality transformation using the $SU(2)$ that acts on $S^2_2$. The sigma model describing the propagation of a string on the NS-NS sector of this background can be cast in the general form 
\begin{equation}
\label{originalL}
L=G_{\mu\nu}(X)\partial_+ X^\mu\partial_- X^\nu+G_{\mu i}(X)(\partial_+ X^\mu L_-^i+\partial_- X^\mu L^i_+)+g_{ij}(X)L^i_+ L^j_-
\end{equation}
Here $L^i_{\pm}=-i {\rm Tr}(t^i g^{-1}\partial_\pm g)$, with $g$ the element of $SU(2)$ parameterized by  $(\psi,\theta_2,\phi_2)$, $t^i$ stand for the generators of $SU(2)$, and
$X^\mu$ run over $AdS_4$, $\zeta$ and $S^2_1$. 
Using the invariance of the sigma model  under $g\rightarrow hg$, with $h\in SU(2)$, the following non-Abelian T-dual NS-NS background can be generated (see for instance \cite{Sfetsos:1996pm})
\begin{equation}
\label{dualL}
{\tilde L}=G_{\mu\nu}\partial_+ X^\mu \partial_- X^\nu+(\partial_+ v_i +\partial_+ X^\mu G_{\mu i})M^{-1}_{ij}(\partial_- v_j-G_{\mu j}\partial_- X^\mu)\, .
\end{equation}
Here $M=g+f$, with $f$ the $SU(2)$ structure constants in the normalization $[t^i,t^j]=i  \epsilon_{ijk}t^k$, and $g$ has been replaced by the Lagrange multipliers $v_i$, $i=1,2,3$, which take values on the Lie algebra of $SU(2)$. The Lagrange multipliers enforce the flat connection condition in the proof of equivalence between the original Lagrangian (\ref{originalL}) and its NATD (\ref{dualL})  \cite{de la Ossa:1992vc}.
The dual metric and NS-NS 2-form read
\begin{eqnarray}
&&{\tilde g}_{ij}=\frac12 M^{-1}_{(ij)}\, , \qquad {\tilde B}_{ij}=\frac12 M^{-1}_{[ij]}\, ,  \qquad
{\tilde G}_{i\mu}=-\frac12 M^{-1}_{[ij]}G_{j\mu}\nonumber\\
&&{\tilde B}_{i\mu}=-\frac12 M^{-1}_{(ij)}G_{j\mu}\, , \qquad
{\tilde G}_{\mu\nu}=G_{\mu\nu}-M^{-1}_{ij}G_{\mu i}G_{\nu j}
\end{eqnarray}
where in our conventions $M_{(ij)}=M_{ij}+M_{ji}$ and $M_{[ij]}=M_{ij}-M_{ji}$.
The dilaton in turn transforms as
\begin{equation}
\phi\rightarrow \phi-\frac12 \log{({\rm det}M)}\, .
\end{equation}

The general procedure to generate the dual RR fields was worked out in  \cite{Sfetsos:2010uq}, and later applied in  \cite{Itsios:2012dc} to obtain general formulas for the duals of 
$M_7\times S^3$ backgrounds with warpings  depending on the $M_7$ directions and RR fluxes consistent with this structure. For backgrounds such as the $AdS_4\times \mathbb{CP}^3$ spacetime in this paper, which contain different warpings for the fibre and the $S^2$ directions of the $S^3$, the transformations of the RR fields were worked out in \cite{Itsios:2013wd}. They are listed in the Appendix for completeness.


Particularizing the previous general results to the $AdS_4\times \mathbb{CP}^3$ background we find for the NS-NS sector
\begin{equation}
\label{metric1}
d{\tilde s}^2=\frac{L^2}{4} ds^2_{AdS_4}+L^2\Bigl(d\zeta^2+\frac14 \cos^2{\zeta}\,(d\theta_1^2+\sin^2{\theta_1}d\phi_1^2)\Bigr)+ds^2_3\, ,
\end{equation}
where $ds^2_3$ stands for the 3-dimensional metric: 
\begin{eqnarray}
\label{dualmetric}
ds^2_3&=&\frac{1}{16\,  \det M}\Bigl[
L^4  \sin^4{\zeta} \Bigl(dr^2+r^2 d\chi^2-\sin^2{\zeta}\, (\sin{\chi}dr+r\cos{\chi}d\chi)^2+\nonumber\\
&&+r^2\cos^2{\zeta}\sin^2{\chi}\, (d\xi+\cos{\theta_1}d\phi_1)^2\Bigr)+  16 r^2dr^2\Bigr]\, ,
\end{eqnarray}
and we have used spherical coordinates to parameterize the $v_i$ Lagrange multipliers living in the Lie algebra of $SU(2)$ 
\begin{equation}
v_1=r\cos{\xi}\sin{\chi}\, , \qquad v_2=r\sin{\xi}\sin{\chi}\, , \qquad v_3=r\cos{\chi}\, .
\end{equation}
In (\ref{dualmetric}) ${\rm det}M$ is given by:
\begin{equation}
 \det M=\frac{L^2}{64} \sin^2{\zeta} \Bigl(16  r^2(\sin^2{\chi}+ \cos^2{\chi}\cos^2{\zeta})+ L^4  \sin^4{\zeta}\cos^2{\zeta}\Bigr)\, .
\end{equation}
The dilaton reads in turn
\begin{equation}
\label{dilaton}
e^\phi=\frac{L}{k}\frac{1}{\sqrt{{\rm det}M}}\, .
\end{equation}
Both the scalar curvature and dilaton remain small when $L\gg 1 \Leftrightarrow N\gg k$. This sets the regime of validity of the NATD supergravity background. 

Moreover, a $B_2$ field is generated that reads:
\begin{eqnarray}
&&B_2=\frac{L^2  \sin^2{\zeta}}{64\, {\rm det}M}\Bigl[ -L^4 r \cos^2{\zeta}\sin^4{\zeta}\cos{\theta_1}\sin{\chi}\, d\phi_1\wedge d\chi-\nonumber\\
&&-16\,  r^2 \, \Bigl( r (\cos^2{\zeta}\cos^2{\chi}+\sin^2{\chi})\, {\rm Vol}({\tilde S}^2)+\sin^2{\zeta}\sin^2{\chi}\cos{\chi}\, d\xi\wedge dr\Bigr) - \nonumber\\
&&-\cos^2{\zeta}\cos{\theta_1}\cos{\chi}\,(L^4  \sin^4{\zeta}+16  r^2)\,  dr\wedge d\phi_1\Bigr]\, ,
\end{eqnarray}
where we have denoted by ${\tilde S}^2$  the 2-sphere spanned by the dual coordinates $(\chi, \xi)$.

Together with this we find the RR sector:
\begin{equation}
\label{RR1}
F_1=\frac{k}{2}   \Bigl(r   \sin^2{\zeta} \sin{\chi}\,  d\chi - \sin^2{\zeta} \cos{\chi}\, dr- r   \sin {2\zeta}  \cos{\chi}\, d{\zeta}\Bigr)
\end{equation}
\begin{eqnarray}
\label{RR2}
{\hat F}_3&=&-\frac{3}{128} k   L^4  \sin^3{2\zeta}  \,d\zeta\wedge 
{\rm Vol}(S^2_1)
+\frac{k}{2}  \Bigl(r \, dr \wedge ( \cos ^2{\zeta} \,  {\rm Vol}(S^2_1)+ \\
&&+ \sin{2\zeta}  \cos{\theta_1}\, d \zeta  \wedge d \phi_1+  \sin{2\zeta}  \sin ^2{\chi} \,d\zeta \wedge d\xi)- r^2 \sin{2\zeta}  \cos{\chi}\, d\zeta \wedge {\rm Vol} ({\tilde S}^2)  \Bigr) \nonumber
\end{eqnarray}
\begin{eqnarray}
\label{RR3}
{\hat F}_5&=&-\frac{1}{64} k L^6   \sin^3{\zeta} \cos{\zeta} \, d{\rm Vol}(AdS_4) \wedge d\zeta+\frac{3}{8} k  L^2   r  \, d{\rm Vol}(AdS_4)\wedge dr+\nonumber\\
&&+\frac{k}{2}     r^2  \Bigl(\cos ^2{\zeta}\, {\rm Vol}(S^2_1)
 +\sin{2\zeta}  \cos{\theta_1}  \, d\zeta  \wedge d \phi_1\Bigr) \wedge dr \wedge {\rm Vol}({\tilde S}^2)\, ,
\end{eqnarray}
where $F_p=d\, C_{p-1}-H_3\wedge C_{p-3}$ and ${\hat F}=F e^{-B_2}$ (see e.g. \cite{Benini:2007kg}). The Page charges will be computed from ${\hat F}$ according to $d * {\hat F}= * j^{Page}$. The electric-magnetic duals of these field strengths can be found in the Appendix.

As we can see, the original $SU(2)$ isometry used in the dualization has disappeared in the dual background, as is usual under a NATD transformation. Moreover, a singularity appears at the fixed point $\zeta=0$, where the squashed $S^3$ used in the dualization shrinks to zero size. This is completely analogous to the singularity that comes out after Abelian T-duality on a shrinking circle. 
At $\phi_1={\rm constant}$  the geometry spanned by $\zeta$ and ${\tilde S}^2$ close to $\zeta=0$ is conformally a singular cone at $\zeta=0$, with an ${\tilde S}^2$ boundary.
Some configurations in the dual CFT such as  some color branes and 't Hooft monopoles will be realized in the gravity dual as branes wrapping this 2-sphere. 

Moreover, at $\zeta=0$, $\phi_1={\rm constant}$, the $B_2$ field reads
\begin{equation}
 B_2=-r\, {\rm Vol}\, ({\tilde S}^2)\, .
 \end{equation}
Large gauge transformations 
\begin{equation}
\label{largegt}
B_2\rightarrow B_2+n\pi {\rm Vol}({\tilde S}^2)
\end{equation}
must then be defined such that the $B_2$ field holonomy around ${\tilde S}^2$ satisfies
\begin{equation}
\label{b}
b=\frac{1}{4\pi^2}|\int B_2\, | \in [0,1)\, ,
\end{equation}
for $r\in [n\pi, (n+1)\pi)$. This division of $\mathbb{R}^+$ in $[n\pi, (n+1)\pi)$ intervals will
have important consequences when we analyze the quantized charges that can be defined in the background.

Indeed, in the absence of any global information  under non-Abelian T-duality, the Lagrange multipliers are just constrained to live in the Lie algebra of $SU(2)$, which implies that $r\in \mathbb{R}^+$. If this variable had been generated through Abelian T-duality, its periodicity would have been fixed by imposing that the gauging procedure used to construct the dual worked as well in non-trivial worldsheets with non-contractible loops \cite{Rocek:1991ps}.  
Extending the non-Abelian T-duality transformation beyond spherical wordsheets is however a long-standing open problem \cite{Alvarez:1993qi}, that brings along our lack of knowledge on the global aspects of the newly generated background. Previous approaches in the literature in order to extract global information  in NATD generated backgrounds have focused on demanding consistency to the associated dual CFT (see for instance \cite{Lozano:2013oma}). Indeed, from  the dual background alone it is not possible to extract any global information, since it is perfectly regular for all $r\in \mathbb{R}^+$. As emphasized in \cite{Lozano:2013oma} this poses a puzzle to the dual CFT, since $r$ should be compact in order to avoid a continuous spectrum of fluctuations.

In our example in this paper the non-trivial relation between the parameter that labels large gauge transformations, $n$, and the $r$ direction, imposed by the condition (\ref{b}), provides crucial global information. We will see that as $r$ approaches infinity in $ [n\pi, (n+1)\pi)$ intervals, the gauge groups of the dual CFT undergo a spectral flow very reminiscent of the 
cascade of \cite{Klebanov:2000hb}, with the important difference that in this case the flow parameter is not the energy scale but the $r$ internal direction. This phenomenon will allow us to relate the different field theories used to describe the solution as $r$ changes interval, to the field theory for $r\in [0,\pi)$.

\subsection{Quantized charges}

Let us start by analyzing the different charges that can be defined in the dual background.

The transformations of the RR fields under non-Abelian T-duality (see Appendix A) show that  a $F_q$ RR field strength transverse to the $S^3$ on which the dualization is performed gives rise to two, $F_{q+1}$ and $F_{q+3}$, field strengths, whose extra components lie on $M_1$ (the space spanned by the $r$-direction) and $M_1\times {\tilde S}^2$, respectively. In turn, a $F_q$ field strength with components along the $S^3$ gives rise to two, $F_{q-3}$ and $F_{q-1}$, RR field strengths in which the $S^3$ directions have either disappeared completely or been replaced by the two directions along ${\tilde S}^2$.
Accordingly, $Dp$-branes transverse to the $S^3$ give rise to $D(p+1)$-branes wrapped on 
$M_1$, and $D(p+3)$-branes wrapped on $M_1\times {\tilde S}^2$. In turn,  $Dp$-branes wrapped on the $S^3$ give rise to $D(p-3)$-branes transverse to $M_1\times {\tilde S}^2$, and $D(p-1)$-branes wrapped on ${\tilde S}^2$ and transverse to $M_1$. As we will see, these branes have a direct interpretation in the dual CFT.

The picture that arises is that a given brane in the original theory gets mapped into a pair of D-branes with relative co-dimension 2. The charges of the original CFT are therefore generically doubled in the dual CFT.  Given that the dual background contains a non-vanishing $B_2$ field, the charges that should be quantized are the Page charges, associated to the currents $d*{\hat F}=*j^{Page}$, with
$\hat{F}=F e^{-B_2}$.
In generic backgrounds, such as the one considered in this paper,
other lower dimensional fluxes besides the ones mentioned above may also be generated through the NATD transformation (such as the $F_1$ in the previous section). Although these fields contribute to the previous Page charges, they do not seem  to give rise to new ones with a direct interpretation in the dual CFT.
In general grounds, given our lack of knowledge of the global properties of the NATD background it is not easy to identify the cycles that should be responsible for these charges.

Let us now particularize this discussion to the $AdS_4\times \mathbb{CP}^3$ background and its NATD.
The $F_6$ flux on the $\mathbb{CP}^3$ gives rise to $F_3$ and $F_5$ field strengths lying respectively on the $\zeta$, $S^2_1$ and $\zeta$, $S^2_1$, ${\tilde S}^2$ directions of the dual geometry. ${\hat F}_5$ is however only non-vanishing  in the presence of large gauge transformations of the $B_2$ field. Since the charges associated to ${\hat F}_3$ and ${\hat F}_5$ are the ones to be interpreted as the ranks of the dual gauge groups, we seem to find two gauge groups associated to each $U(N)$ in the original theory, when  we allow for large gauge transformations in the dual background.  We will see that consistently with this picture, two types of color branes and baryon vertices exist in the dual theory, with tadpoles, for the last, given by the quantized charges associated to these fields.

The $F_2$ flux of the $AdS_4\times \mathbb{CP}^3$  background transforms in turn in
a well-defined manner under non-Abelian T-duality  when it lies on the $S^2$ that remains neutral under the dualization. This is the $S^2$ seated at $\zeta=0$, which we have previously referred to as $S_1^2$. $F_2$  
gives then rise to both ${\hat F}_3$ and  ${\hat F}_5$ field strengths in the dual, lying on the $M_1$, $S^2_1$ and $M_1$, $S^2_1$, ${\tilde S}^2$ directions, respectively. Since the associated quantized charges are to be interpreted as the levels of the dual quiver,  D3 and D5  't Hooft monopoles should exist in the dual geometry with tadpoles given by these charges. These branes are constructed in the next section. 

The Page charges associated to the previous field strengths are obtained from the quantization condition
\begin{equation}
\label{quantcond}
\frac{1}{2\, \kappa_{10}^2}\int {\hat F}_p=T_{8-p}\, N_{8-p}\, .
\end{equation}
Integration over the ($\zeta, \theta_1, \phi_1$) components of ${\hat F}_3$ gives a charge: 
\begin{equation}
N_5=\frac{kL^4}{64\pi}\, ,
\end{equation}
that should be interpreted in the dual CFT as the rank of one of  the dual gauge groups.
Note that this charge cannot be an integer if $L$ satisfies (\ref{rank}). Instead, $L$ will be defined in the dual theory from the ranks and levels of the dual gauge groups through new relations that we will derive below. That integer charges are mapped onto non-integer ones is a generic feature under non-Abelian T-duality, and it is due to the fact  that the volume of the $n$-dimensional space on which the dualization is performed  violates the condition
\begin{equation}
\label{tensions}
T_{p-n}=  (2\pi)^n T_p\, ,
\end{equation}
that is implicit in (\ref{quantcond}). In our particular case the tensions of the D2 color branes of the original background and the D5 color branes of the NATD are related through the volume of the $S^3$:  $T_2=2\pi^2\, T_5$, which indeed violates (\ref{tensions}). 

As we already mentioned, we can see from (\ref{RR3}) that the $\zeta, S^2_1, {\tilde S}^2$ components of ${\hat F}_5$ are zero in the dual background. Therefore in the absence of large gauge transformations the rank of the second gauge group vanishes. This is not so however in the presence of large gauge transformations, where ${\hat F}_5$ becomes  
\begin{equation}
\label{F3F5}
{\hat F}_5\rightarrow  {\hat F}_5-{\hat F}_3\wedge n\pi {\rm Vol} ({\tilde S}^2)\, ,
\end{equation}
and D3-brane charge, $N_3=n N_5$, is generated through (\ref{quantcond}).

The levels of the dual theory are in turn given by
\begin{equation}
\label{k3k5}
k_5=\frac{1}{(2\pi)^2}\int {\hat F}_3\, , \qquad k_3=\frac{1}{(2\pi)^4}\int  {\hat F}_5\, ,
\end{equation}
with the integration taking place over  $(r,\theta_1,\phi_1)$ for ${\hat F}_3$ and $(r,\theta_1,\phi_1)$ plus
${\tilde S}^2$ for ${\hat F}_5$. Note that in the presence of large gauge transformations
\begin{equation}
\label{k3mod}
k_3=\frac{1}{(2\pi)^4}\int  {\hat F}_5\rightarrow
\frac{1}{(2\pi)^4}\int ( {\hat F}_5-{\hat F}_3 \wedge n\pi {\rm Vol}({\tilde S}^2))
\end{equation}
and $r$ must be integrated on $[n\pi, (n+1)\pi)$. This guarantees positive levels for all $n$: 
\begin{equation}
\label{kvalues}
k_5= k\, \frac{(2n+1)\pi}{4}\, , \quad k_3=k\, \frac{(3n+2)\pi}{12}\, .
\end{equation}
The levels of the two dual gauge groups thus satisfy $(3n+2)k_5=3(2n+1)k_3$ , showing that they can be integers at the same time. In the absence of large gauge transformations $2k_5=3k_3$, and both levels are non-zero even if the rank of one of the dual gauge groups is fully depleted. The expressions in (\ref{kvalues}) show on the other hand that $k_3$ and $k_5$ cannot be integers if the original level $k$ is an integer. 
This is again due to the fact that the relation between the tensions given by (\ref{tensions}) is violated by the non-Abelian T-duality transformation. 

The NATD background is fully consistent globally if $L$ satisfies, in terms of the quantized charges of the dual background,
\begin{equation}
\label{k5}
k_5\, L^4=32 \pi^2 (N_3+\frac{N_5}{2})\, , \qquad 
\end{equation}
or, 
\begin{equation}
\label{k3}
k_3\, L^4=16\pi^2 (N_3+\frac23 N_5)\, .
\end{equation}
These expressions resemble the functional dependence of the radius of curvature of the original ABJM theory
on the 't Hooft coupling, $\lambda=N/k$, 
\begin{equation}
k\, L^4=32\pi^2 N\, ,
\end{equation}
and are very suggestive of a 't Hooft coupling defined as the quotient of a rank by a level in the dual CFT. We will come back to this point in section 4.3. 

Finally, the regime of validity of the NATD solution becomes, in terms of the dual ranks and levels, 
$N_5, N_3\gg k_5, k_3$.

\subsection{Supersymmetry}

The condition that SUSY is preserved under non-Abelian T-duality is believed to be given by the vanishing of the Kosmann derivative \cite{Sfetsos:2010uq}.  Indeed it was proven that this is the case when performing an $SU(2)$ isometry T-duality on a geometry with an $SO(4)$ isometry in \cite{Itsios:2012dc}. Here we will take the view that this is also the case when the isometry is $SU(2)$, and leave a detailed proof for work in preparation \cite{ENY}.

The Kosmann derivative is defined as \cite{Kosmann1972}
\beq
\mathcal{L}_{k}\epsilon=k^{a}D_{a}\epsilon+\frac{1}{8}\left(dK\right)_{ab}\Gamma^{ab}\epsilon
\eeq
where $k= k^a \partial_a$ are Killing vectors and $K=k_adx^a$ are their dual one forms. For our $SU(2)$ isometry there are 3 Killing vectors
\begin{align}
k_{(1)}&= -\cos\phi_2\partial_{\theta_2}+\cot\theta_2 \sin\phi_2\partial_{\phi_2}- \csc\theta_2\sin\phi_2\partial_{\psi},\\
k_{(2)}&= -\sin\phi_2 \partial_{\theta}-\cot\theta_2 \cos\phi_2 \partial_{\phi_2}+\csc\theta_2\cos\phi_2\partial_{\psi}\\
k_{(3)}&= -\partial_{\phi_2}
\end{align}
which satisfy the $SU(2)$  conditions $[k_{(a)},~k_{(b)}]=-\epsilon_{abc} k_{(c)}$. The Kosmann derivative must vanish for each of these for SUSY to be preserved. Any additional conditions that the spinor must satisfy in order to make this so will reduce the supersymmetry, typically (but not always) by half.

The Killing spinor of ABJM is rather complicated and we defer an explicit derivation until Appendix \ref{sec: AB}. However one can make use of the fact that the geometry of ABJM lends itself to a 4+6 split to express the spinor as
\beq
\epsilon = \chi_{AdS_4}\otimes \eta_{\mathbb{CP}^3}.
\eeq
The 4d spinor $\chi_{AdS_4}$ preserves maximal supersymmetry, i.e. 4 supercharges, while the 6d spinor $\eta_{\mathbb{CP}^3}$ preserves 6. This gives a total of 24 preserved by $\epsilon$. The specific form of $\eta_{\mathbb{CP}^3}$ is given by (\ref{CP3spinors}), and in general depends on all coordinates of $\mathbb{CP}^3$ as well as 6 arbitrary constants. However it is possible to define a restricted spinor that depends only on $\zeta$ and two constants. In the frame of eq (\ref{eq:frame}) this is given by
\beq\label{spinorkosmann}
\tilde{\eta}_{\mathbb{CP}^3}=(0,c_1\cos\zeta,c_1\sin\zeta,0,0,-c_2\sin\zeta,c_2\cos\zeta,0)^{T}.
\eeq
Such a spinor represents 2 preserved supercharges, but is constant with respect to the directions $\theta_2,\phi_2,\psi$, so we need only to satisfy the condition 
\beq
(k^{a}\omega_{a,bc}+\frac{1}{2}\left(dK\right)_{bc})\gamma^{bc}\tilde{\eta}_{\mathbb{CP}^3}=0
\eeq
where $\gamma^a$ are 6 dimensional gamma matrices (see appendix \ref{sec: AB}). It is a relatively short computation to show that
\begin{align}
dK_{(1)} &=\frac{1}{2}\bigg(2\cos\zeta\big(\cos\theta_2\cos\psi\sin\phi_2+\cos\phi_2\sin\psi\big)e^{14}+\sin\theta_2\sin\phi_2\big(e^{23}+3e^{45}-\nn\\
&~~~~\cos2\zeta(2e^{16}+e^{23}+e^{45})\big)+2\cos\zeta\cos\theta_2\sin\phi_2\big(\sin\psi(e^{15}+e^{46})-\cos\psi e^{56}\big)-\nn\\
&~~~~2\cos\zeta\cos\phi_2\big(\cos\psi(e^{15}+e^{46})+\sin\psi e^{56}\big)\bigg),\nn\\
dK_{(2)} &=\frac{1}{2}\bigg(\cos\phi_2\sin\theta_2\big(-e^{23}-3e^{45}+\cos2\zeta(2e^{16}+e^{23}+e^{45})\big)-\nn\\
&~~~~2\cos\zeta\cos\psi\big(\sin\phi_2(e^{15}+e^{46})+\cos\theta_2\cos\phi_2(e^{14}-e^{56})\big)+\nn\\
&~~~~2\cos\zeta\sin\psi\big(-\cos\theta_2\cos\phi_2(e^{15}+e^{49})+\sin\phi_2(e^{14}-e^{56})\big)\bigg),\nn\\
dK_{(3)} &=\frac{1}{2} \cos\theta_2 \big(e^{23}+3e^{45}-\cos2\zeta(2e^{16}+e^{23}+e^{45})\big)-\nn\\
&~~~~\cos\zeta\sin\theta_2\big(\sin\psi(e^{15}+e^{46})+\cos\psi(e^{14}-e^{56})\big),
\end{align}
and using eq (\ref{eq:spinconnection}) one finds that 
\beq
dK_{(i)} + k^{a}_{(i)}\omega_{a,bc}e^{bc}=0.
\eeq
This proves that the Kosmann derivative vanishes for any spinor that is constant with respect to $\theta_2,\phi_2,\psi$, and in particular for the one given by eq (\ref{spinorkosmann}). We can therefore conclude that the internal space of the Type IIB geometry preserves at least 2 supercharges, which gives a total of 8 supercharges preserved by the whole space. That is, the NATD solution preserves at least $\mathcal{N}=2$ supersymmetries in 3 dimensions, with possible enhancements for specific values of $k$ as in ABJM, or at specific points in the space.
Note that $\mathcal{N}=2$ is fully consistent with the isometries present in the dual background.

\section{Analysis of the dual CFT}

In this section we identify various branes associated to the quantized charges found in section 3. We show that large gauge transformations generate changes in these charges 
and thus in the field theory used to describe the solution.
We find that in parallel  with ABJM  there are particle-like branes whose tadpoles and conformal dimensions are associated to these charges. Putting all this information together we finally conjecture a $CFT_3$ dual to the new $AdS_4$ solution. This
field theory changes as $r$ moves from 0 to infinity through a phenomenon very reminiscent of the cascade \cite{Klebanov:2000hb} that allows for a possible interpretation in the CFT of the running of the non-compact direction.

\subsection{Color branes}

Let us start by studying the supersymmetric locii of probe branes filling $\mathbb{R}_{1,2}$. According to the discussion in section 3.1 we expect to find  D3-branes wrapped on $\mathbb{R}_{1,2}\times M_1$ and D5-branes wrapped on $\mathbb{R}_{1,2}\times M_1\times {\tilde S}^2$. These branes should be responsible for the $N_3$ and $N_5$ charges.

The DBI action of a D3-brane wrapped on $\mathbb{R}_{1,2}\times M_1$ reads, at $\zeta=0$:
\begin{equation}
\label{D3DBI}
S_{DBI}^{D3}=-T_3 \int e^{-\phi} \sqrt{g}=-T_3\int \frac{kL^2}{8} r\, \rho^3=-\frac{k_5 L^2}{32 \pi^2}\int \rho^3\, .
\end{equation}
This expression is equal and of opposite sign to  the CS action, which implies that the D3-branes are BPS when located at $\zeta=0$ \footnote{Note that this is similar to the findings in \cite{Imeroni:2008cr} for color D2-branes in the $\beta$-deformed ABJM background, which preserves $\mathcal{N}=2$ supersymmetries.}, the tip of the cone with ${\tilde S}^2$ boundary discussed in section 3. (\ref{D3DBI}) is, on the other hand, identical to the BI action of the color D2-branes of the original ABJM theory, with $k$ replaced by $k_5$. The level of the probe D3-brane is thus given by the Page charge $k_5$. This is confirmed by the 
analysis of the quadratic fluctuations of the D3-brane, which gives an effective YM coupling
\begin{equation}
S=\int \frac{1}{g_{D3}^2} F^2_{\mu\nu} \qquad {\rm with}\qquad  \frac{1}{g_{D3}^2}=\frac{k_5}{8}\rho\, .
\end{equation}

A D5-brane wrapped on $\mathbb{R}_{1,2}\times M_1\times {\tilde S}^2$  also experiences a no-force condition when located at $\zeta=0$. This brane should be associated to the gauge group with rank $N_5$.
In this case  the DBI action and (minus)  the CS action read\footnote{For the computation of the CS action we have used that 
$d(C_6-B_2\wedge C_4)=F_7-B_2\wedge F_5$ in our background, 
and that $F_7$ vanishes at $\zeta=0$.},
\begin{equation}
\label{DBID5}
S_{DBI}^{D5}=-T_5 \int e^{-\phi} \sqrt{|{\rm det}(g-B_2)|}=-T_5\int 
\frac{kL^2}{2}\pi \Bigl(r-n\pi\Bigr)r\rho^3=-\frac{k_3 L^2}{32 \pi^2}\int \rho^3\, .
\end{equation}
 The quadratic fluctuations give in turn an effective YM coupling:
\begin{equation}
S=\int \frac{1}{g_{D5}^2} F^2_{\mu\nu} \qquad {\rm with}\qquad  \frac{1}{g_{D5}^2}=\frac{k_3}{8}\rho\, .
\end{equation}
These results identify $k_3$ as the level of the gauge group associated to D5-branes. Note that $k_3$, and therefore $g_{D5}^2$, is positive for $r\in [n\pi, (n+1)\pi)$ (which is implicit in the integration in (\ref{DBID5})), while already for $r\in [(n-1)\pi,n\pi)$ it becomes negative. In that interval it can be made positive  if  $n\rightarrow n-1$. 
Moreover, the D5-brane CS coupling
$S^{D5}_{CS}=-T_5\int C_4\wedge B_2$,  includes a term
\begin{equation}
S^{D5}_{CS}=  -T_5\int C_4\wedge n\pi {\rm Vol}({\tilde S}^2)=-n T_3 \int C_4\, ,
\end{equation}
which
implies that  the D3-brane charge dissolved in the D5-brane decreases in  one unit when moving from the $[n\pi, (n+1)\pi)$ interval to the previous one.  Since, as we have seen, $N_3=nN_5$, this changes the Page charge of the D3-branes as:
\begin{equation}
N_3\rightarrow N_3-N_5\, .
\end{equation}
This resembles Seiberg duality in $\mathcal{N}=1$  \cite{Seiberg:1994pq}.
Note moreover that  both the D3 and D5 color branes are located at the tip of the cone, with the D5-branes wrapping a collapsing 2-cycle at the singularity, and acting as domain walls in $AdS_4\times M_1$. These branes thus seem to play the role of the fractional branes in  \cite{Gubser:1998fp}, with the important difference that  both the D3
and the D5-branes  are now lying on the $r$ internal direction.
We will discuss further our interpretation of the $N_3\rightarrow N_3-N_5$ flow in section 4.3.

Finally, D5-branes wrapped on $\mathbb{R}_{1,2}\times M_1\times {\tilde S}^2$ can be shown to arise from systems of coincident $(m+1)$ D3-branes wrapped on  $\mathbb{R}_{1,2}\times M_1$ by Myers dielectric effect \cite{Myers:1999ps}.
This holds true even in the absence of large gauge transformations. Note that the D3-branes are BPS at $\zeta=0$, where a non-trivial 2-sphere ${\tilde S}^2$ exists, into which they can expand. The relevant dielectric coupling in the CS action responsible for this expansion is
\begin{equation}
S_{diel}^{D3}=i\,\frac{T_3}{2\pi}\int {\rm STr}\{(i_X i_X)(C_6-C_4\wedge B_2)\}\, ,
\end{equation}
which gives:
\begin{equation}
S_{diel}^{D3}
=\frac{k_3 L^2}{32\pi^2}\frac{m+1}{\sqrt{m(m+2)}}\int \rho^3\, .
\end{equation}
This reproduces the CS action for a D5-brane wrapped on $\mathbb{R}_{1,2}\times M_1\times {\tilde S}^2$, given by minus (\ref{DBID5}),
 in the large $m$ limit, and coincides exactly with minus the dielectric contribution to the DBI action of $m+1$ coincident D3-branes, computed from
\begin{equation}
S_{DBI}=-T_3\int {\rm STr} \{e^{-\phi}\sqrt{g}\sqrt{{\rm det}Q}\}\, ,
\end{equation}
where 
\begin{equation}
Q^i_j=\delta^i_j+\frac{i}{2\pi}[X^i, X^k](g-B_2)_{kj}\, .
\end{equation}
This result shows that the dipole coupling in the action  of coincident $(m+1)$ D3-branes reproduces a D5-brane wrapped on ${\tilde S}^2$. The monopole coupling gives in turn $(m+1)$ D3-branes, described by (\ref{D3DBI}).

The existence of two types of BPS branes  with an interpretation as color branes supports our statement in the previous section for a doubling of each of the 
$U(N)$ gauge groups of the ABJM 2-node quiver. Moreover, the role played by the D5-branes as ``fractional'' branes, further suggests $N_3+N_5$ and $N_3$ as the ranks of the respective gauge groups.  
We will see below that di-monopole, 't Hooft monopole, di-baryon and baryon vertex configurations can be constructed  for all  $N_3, k_5$, $N_5, k_3$ charges. BPS D3-branes wrapped on $\mathbb{R}_{1,2}\times \zeta$ as well as D5-branes wrapped on $\mathbb{R}_{1,2}\times \zeta\times {\tilde S}^2$, dual to the BPS D6-branes wrapped on $\mathbb{R}_{1,2}\times \zeta\times S^3$ of the original ABJM theory, can also be constructed.

\subsection{Particle-like branes}

Particle-like branes in ABJM are topologically non-trivial configurations  associated to  the homology cycles of the $\mathbb{CP}^3$ internal space \cite{Aharony:2008ug}. A D2-brane wrapped on an $S^2$ captures the RR 2-form flux, and develops a tadpole with charge equal to the level of the dual gauge group. This brane is interpreted in the field theory as a 't Hooft monopole. A D4-brane wrapped on a $CP^2$ does not capture any of the background fluxes, and has the same baryon charge and dimension as the di-baryon operator of the field theory \cite{Gubser:1998fp,Berenstein:2002ke}. The D6-brane wrapped on the $CP^3$ captures the flux of the $F_6$ field, and develops a tadpole with charge equal to the rank of the dual gauge group. This brane is the analogue of the baryon vertex in $AdS_5\times S^5$ \cite{Witten:1998xy}.  Finally, a D0-brane is gauge invariant and has conformal dimension equal to the level of the gauge group. This brane is  referred as the di-monopole.

In this section we show that all these brane configurations exist in the NATD theory associated to the $N_3, k_5$, $N_5, k_3$ charges found in section 3.1.

\subsubsection{Di-monopoles}
 
Di-monopoles in ABJM have conformal dimension equal to the level of the gauge group and correspond to Wilson lines in the $({\rm Sym}_k, \bar{\rm Sym}_k)$ representation \cite{Gaiotto:2009mv}. In the gravity dual they are realized as D0-branes, with mass $m_{D0}L=k$ and no tadpoles. In the NATD background there are two branes dual to the di-monopole configuration, namely a D1-brane wrapped on $M_1$ and a D3-brane wrapped on $M_1\times {\tilde S}^2$. 
As for the color branes discussed in the previous section the D3-brane arises from a system of coincident D1-branes by Myers dielectric effect. 

The D1-branes have an energy proportional to $k_5$ when they sit at $\zeta=0$
\begin{equation}
S_{DBI}^{D1}=-T_1 \int e^{-\phi} \sqrt{g}=-\frac{k_5}{L}\int \sqrt{|g_{tt}|}\, dt
\end{equation} 
while the D3-branes have an energy proportional to $k_3$ also while seated at $\zeta=0$
\begin{equation}
S_{DBI}^{D3}=-T_3 \int e^{-\phi} \sqrt{|{\rm det}(g-B_2)|}=-\frac{k_3}{L}\int \sqrt{|g_{tt}|}\, dt\, ,
\end{equation}
with $k_5$ and $k_3$ given by (\ref{kvalues}). We thus find that there are di-monopole configurations associated to the two charges that can be interpreted as levels in the dual CFT. These (or combinations thereof)\footnote{See the discussion in section 4.3.} should correspond to
 the conformal dimensions of the di-monopole operators in the dual CFT.

\subsubsection{'t Hooft monopoles}

 't Hooft monopoles are realized in ABJM as D2-branes wrapping 2-spheres in the $\mathbb{CP}^3$. These branes have  tadpoles of $k$ units and correspond to
$k$ Wilson lines in the representation ${\rm Sym}_k$ ending on a point  \cite{Aharony:2008ug, Gaiotto:2009mv}.  Of these, the D2-brane wrapped on the $S^2_1$ at $\zeta=0$ 
survives after the NATD transformation.
According to our general discussion, we should find  a D3-brane wrapped on $M_1\times S^2_1$, and a D5-brane wrapped on $M_1\times S^2_1\times {\tilde S}^2$ in the NATD. We have indeed found these branes, with tadpoles with $k_5$ and $k_3$ charges, respectively.
These should correspond to this number of Wilson lines in the ${\rm Sym}_{k_5}$, ${\rm Sym}_{k_3}$ representations ending on a point. As in the previous sections, the D5-brane arises after the expansion of coincident D3-branes into the ${\tilde S}^2$  through Myers dielectric effect. 

From (\ref{RR2}) we see that a D3-brane wrapped on $M_1\times S^2_1$  carries a tadpole with charge
\begin{equation}
S^{D3}_{CS}=-2\pi \,T_3\int {\hat F}_3\, A_t=-k_5 \int dt \, A_t\, ,
\end{equation}
when located at $\zeta=0$.
In turn, a D5-brane wrapped on $M_1\times S^2_1\times {\tilde S}^2$ carries a tadpole
\begin{equation}
S^{D5}_{CS}=-2\pi\, T_5\int ( {\hat F}_5-{\hat F}_3\wedge B_2)\, A_t=-k_3 \int dt\, A_t\, ,
\end{equation}
when placed at $\zeta=0$. 

Finally,  the $k_3$ tadpole charge of the D5-brane in its dielectric description arises in a system of coincident D3-branes from the dielectric coupling
\begin{equation}
S^{D3}_{diel}=-i\, T_3 \int {\rm STr}\{(i_X i_X)(F_5-F_3\wedge B_2)\}\, A_t\, .
\end{equation}

\subsubsection{Di-baryons}

The di-baryon operator in the ABJM theory, ${\rm det} A=\epsilon_{i_1\dots i_N}\epsilon^{j_1\dots j_N}A^{i_1}_{j_1}\dots A^{i_N}_{j_N}$, with $A$ one of the bi-fundamentals,
is realized in the gravity side as a D4-brane wrapped on  the $(\zeta$, $\psi)$ directions and  any of the two 2-spheres $S^2_1$, $S^2_2$. These sub-manifolds span a $\mathbb{CP}^2$ inside the $\mathbb{CP}^3$. The energy of the D4-brane coincides with the conformal dimension of the di-baryon operator,  $m_{D4}\, L=N$ \cite{Gubser:1998fp, Berenstein:2002ke}.

The di-baryon wrapped on $S^2_2$ survives after the NATD transformation. Thus, according to our general pattern above, we expect to find a D1-brane wrapped on $\zeta$ and a D3-brane wrapped on $\zeta\times {\tilde S}^2$ as duals of this configuration.
 These branes do indeed have conformal weights equal to $N_5$ and $N_3$, respectively, and no tadpoles on their worldvolumes.

The D1 di-baryon sits at $r=0$, where its DBI action is given by
\begin{equation}
S_{DBI}^{D1}=-T_1 \int e^{-\phi} \sqrt{|g_{tt}|\,g_{\zeta\zeta}}=-\frac{N_5}{L}\int \sqrt{|g_{tt}|} \,dt\, .
\end{equation} 
Therefore $m_{D1}\, L=N_5$. A
D3-brane wrapped on ($\zeta, \chi, \xi$) and sitting at $r=0$ has in turn a DBI action
\begin{equation}
S_{DBI}^{D3}=-T_3 \int e^{-\phi} \sqrt{|g_{tt}|\,g_{\zeta\zeta}{\rm det}(g-B_2)|_{{\tilde S}^2}}=-\frac{n\,N_5}{L}\int \sqrt{|g_{tt}|} \,dt= -\frac{N_3}{L}\int \sqrt{|g_{tt}|} \,dt\, ,
\end{equation} 
and $m_{D3}\, L=N_3$. The energy of this di-baryon is  entirely due to the $\bar{D1}$-branes that are induced in its worldvolume by the contribution of large gauge transformations
to the CS term $-T_3 \int C_2\wedge B_2$:
\begin{equation}
S^{D3}_{CS}=-T_3 \int C_2\wedge n\pi {\rm Vol}({\tilde S}^2)=-n T_1 \int C_2\, .
\end{equation}
Therefore it does not exist when the dual gauge group with rank $N_3$ is fully depleted.

The existence of di-baryon configurations in the NATD background suggests the presence of bifundamental fields in the dual CFT. This is also suggested by the giant graviton analysis in section 5.

\subsubsection{Baryon vertices}

The baryon vertex is realized in ABJM as a D6-brane wrapping the whole internal space. This brane has a tadpole of $N$ units and corresponds to $N$ Wilson lines ending on an $\epsilon$ tensor. As before, there are two D-brane configurations dual to it, consisting in this case of a D3-brane wrapped on  $\zeta \times S^2_1$ and a D5-brane wrapped on $\zeta \times S^2_1\times {\tilde S}^2$. These branes have tadpoles with charges equal to $N_5$, $N_3$, respectively, and should correspond to this number of Wilson lines ending on $\epsilon^{i_1\dots i_{N_5}}$ or $\epsilon^{i_1\dots i_{N_3}}$ epsilon tensors \footnote{Note that if the rank of one of the gauge groups is $N_5+N_3$, as we will propose in the next section for the dual CFT describing the solution for $r\in [n\pi, (n+1)\pi)$ for $n\neq 0$, the $N_5+N_3$ Wilson lines ending on $\epsilon^{i_1\dots i_{N_5+N_3}}$ will be realized in the AdS background as bound states of D5 and D3-branes.}.

From (\ref{RR2}) we see that a D3-brane wrapped on $\zeta\times S^2_1$  has a tadpole with charge
\begin{equation}
S^{D3}_{CS}=-2\pi \,T_3\int {\hat F}_3\, A_t=N_5 \int dt \, A_t\, .
\end{equation}
In turn, a D5-brane wrapped on $\zeta\times S^2_1\times {\tilde S}^2$ carries a tadpole
\begin{equation}
S^{D5}_{CS}=-2\pi\, T_5\int {\hat F}_5\, A_t=-n\, N_5 \int dt\, A_t=-N_3 \int dt\, A_t\, .
\end{equation}
Therefore this second baryon vertex only exists in the presence of large gauge transformations.

\subsection{A conjecture for the dual CFT}

Let us now put all these results together to try to propose a $CFT_3$ dual to our new $AdS_4$ solution. 

The existence of two types of color branes suggests that each gauge group of the original $U(N)_k\times U(N)_{-k}$ Chern-Simons theory is replaced by a two node quiver.
This is confirmed by the quantized charges that can be defined in the dual background. 
Consistently, we have also found that the same particle-like brane configurations that exist in ABJM exist in the NATD for each of the two dual gauge groups. 
Note that some of these branes have to sit at the singularity, $\zeta=0$. Still, their worldvolumes are perfectly well-defined.

An important role is played by large gauge transformations of the $B_2$ field, that can be defined thanks to the existence of the non-trivial ${\tilde S}^2$ at $\zeta=0$. For a given large gauge transformation with  parameter $n$, the non-compact direction $r$ must live in the interval $[n\pi, (n+1)\pi)$. The D3 and D5 Page charges that should be related to the ranks of the two dual gauge groups are given in this interval by $N_5$, $N_3=nN_5$, and
 change under $n\rightarrow n-1$ as
\begin{equation}
\label{transN}
N_5\rightarrow N_5\, , \qquad N_3\rightarrow N_3-N_5\, .
\end{equation}
As we have seen, this change allows to have positive squared couplings as $r$ decreases.

Now, if we identify the ranks of the gauge groups of the 2-node quiver as  \cite{Benini:2007gx} 
\begin{equation}
r_l=N_3+N_5\, , \qquad r_s=N_3\, ,
\end{equation}
(\ref{transN}) reduces to Seiberg duality \cite{Seiberg:1994pq}
\begin{equation}
\label{seiberg}
r_l\rightarrow r_s\, , \qquad r_s\rightarrow 2\,r_s-r_l\, .
\end{equation}
Note however that from (\ref{kvalues}) we learn that the levels should transform as well, and quite non-trivially, under $n\rightarrow n-1$:
\begin{equation}
\label{transk}
k_3\rightarrow 3\, k_5-5\, k_3\, , \qquad k_5\rightarrow 7\, k_5-12\, k_3\, ,
\end{equation} 
while they are both non-vanishing for $n=0$, where they satisfy $2k_5=3k_3$. The behavior of the ranks under the flow as well as the results in section 4.2, which associated $k_3$ with the level of the D5-branes and $k_5$ with the level of the D3-branes, suggest a 4-node quiver:
\begin{equation}
\label{quiver}
U(N_3+N_5)_{k_3}\times U(N_3)_{k_5}\times U(N_3+N_5)_{-k_3}\times U(N_3)_{-k_5}\, ,
\end{equation}
with $N_3=nN_5$ and $k_3$ and $k_5$ as given by (\ref{kvalues}), as dual CFT to the solution for $r\in [n\pi, (n+1)\pi)$. Note however that, as in \cite{Benini:2007gx}, there is an ambiguity in the assignment of levels to the gauge groups. Here we have followed the criterium that $k_5$ should be the level associated to color D3-branes, responsible in turn for the rank of the second gauge group. There may be however other arrangements with a more direct physical interpretation\footnote{Notice for instance that $k_5-2k_3$ remains invariant under the flow.}.
Note as well that in the absence of any global information coming from the NATD transformation we can only conjecture that the nature of the gauge groups is not changed under the transformation. Finally, the existence of the di-baryon and baryon vertex configurations discussed in the previous section suggests that the matter content should consist on two sets of 
$N_3+N_5$ and $N_3$ bifundamentals, one for each pair of dual gauge groups with opposite levels.
We will see that this is supported by the giant graviton analysis in the next section.

The previously proposed  CFTs would then provide the best description  (one in which the squared gauge couplings are positive) of the NATD solution as $r$ runs from 0 to infinity in intervals of length $\pi$.
Invariance under large gauge transformations 
suggests, on the other hand, that these theories should have the same field content as $r$ varies, with only the ranks and levels shifting by some integer\footnote{It is puzzling however that the di-baryon configurations found in section 4.2.3 have to sit at $r=0$.}. In other words, we expect the dual CFT to be self-similar under the flow defined by (\ref{transN}) and (\ref{transk}).
This would
imply that all 4-node quivers should be equivalent to, 
in particular, the two node quiver $U(N_5)_{k_3}\times U(N_5)_{-k_3}$, that describes the solution for $r\in [0,\pi)$. This suggests a dual CFT similar to ABJM
albeit with reduced supersymmetry. We will further speculate in the Conclusions with a possible brane realization in Type IIB  based on the way NATD acts on the original ABJM brane system. In this field theory $N_5$ and $k_3$ would be subjected to the relation
\begin{equation}
k_3\, L^4=\frac{32}{3}\pi^2 N_5\, ,
\end{equation}
which is derived from either (\ref{k3}) or (\ref{k5}) for $n=N_3=0$. This is suggestive of a 't Hooft coupling
$\lambda\sim N_5/k_3$ in the dual CFT.

It would of course be interesting to understand better the nature of the flow that is implied by large gauge transformations. Note that given that the gauge couplings are defined in terms of the levels $k_3$ and $k_5$, they must transform quite non-trivially. It would in particular be interesting to see if an interpretation in terms of the Higgsing in  \cite{Aharony:2000pp} could be at work in this case. Finally, note that this is different from 
Seiberg duality in 3d  CS-matter theories  \cite{Giveon:2008zn},  of which Seiberg duality in ABJ \cite{Aharony:2008gk,Aharony:2009fc}, which relates $U(N+l)_k\times U(N)_{-k}$ to $U(N)_k\times U(N+k-l)_{-k}$ $\mathcal{N}=3, 6$ superconformal CS-matter theories, is a particular case.

\section{Giant graviton configurations}

Mesonic operators built out of bifundamental fields correspond in the gravity dual to dual giant gravitons  \cite{Grisaru:2000zn,Hashimoto:2000zp} wrapped on an $S^{n-2}$ inside $AdS_n$ \cite{Corley:2001zk}. 
We show in this section that D3 and D5 dual giant graviton configurations exist in the NATD background that should be dual to these operators in the CFT. More information could be extracted from these configurations upon quantization, following  \cite{Biswas:2006tj}-\cite{Martelli:2006vh} (see also \cite{Bergman:2012qh}). 

In this section it is more convenient to express the $AdS_4$ metric in global coordinates, dual to radial quantization in the CFT
\beq
ds^2_{Ads_4}=-(1+\rho^2)dt^2+ \frac{d\rho^2}{1+\rho^2}+ \rho^2 ds^2_{S^2}.
\eeq


Let us start by analysing the supersymmetric fuzzy sphere solution constructed in \cite{Nishioka:2008ib} in ABJM. This object is the dimensional reduction of the M-theory dual giant graviton, and no longer propagates in Type IIA, but predicts a new class of supersymmetric states in ABJM.
It is described as a D2 brane with embedding $\sigma^i=(t,S^2)$ and non trivial worldvolume flux
\beq\label{eq: D0s}
F= \frac{M}{2} Vol(S^2),
\eeq
dissolving $M$ D0-branes. The action of the D2 brane is then given by
\begin{align}
S_{D2} &= -T_2 \int d^3\sigma\, e^{-\phi} \sqrt{-\det(g+ 2\pi F)}+T_2 \int C_3\nn\\[3 mm]
& = \frac{kL^2}{8\pi}\int dt \left( \rho^3- \sqrt{1+\rho^2}\sqrt{\frac{16M^2 \pi^2}{L^4}+ \rho^4}\right),
\end{align}
where in the first line the pullback onto $\sigma$ should be understood.
As the integrand has no t dependence the minimum energy condition is obtained by minimising $S_{D2}$, which leads to 
\beq
\r= 0, ~~~ \r=\frac{4M \pi}{L^2}.
\eeq
The second of these is the fuzzy sphere solution that descends from the M-theory giant graviton, in either case the energy is given by
\beq
E=\frac{k M}{2}.
\eeq

We seek a similar dielectric brane solution after the T-duality. The established $SU(2)$ transformation rules tell us to seek a D3-brane on $\sigma_1^i=(t,S^2,r)$ and a D5-brane on $\sigma_2^i=(t,S^2,r, \tilde{S}^2)$. In order to do this we will need the fact that, on these embeddings
\begin{align}
d\mathcal{L}_{WZ, D3}&= \hat{F}_4 = d\left(\frac{1}{8} k L^2 \rho^3 r dt\wedge Vol(S^2) \wedge dr\right)\nn \\[3 mm]
d\mathcal{L}_{WZ, D5}&=\hat{F}_7 = d\left(\frac{1}{8} k L^2 \rho^3 r(r-n \pi)dt\wedge  Vol(S^2) \wedge dr\wedge  Vol(\tilde{S}^2)\right),
\end{align}
where in writing the D5 WZ term we have first performed $n$ large gauge transformations.

Let us look at the D3 brane in some detail. We once more turn on a worldvolume gauge field as in eq (\ref{eq: D0s}), which will now be associated to $M$ D1-branes, and fix the coordinates orthogonal to $\sigma_1^i=(t,S^2,r)$ to arbitrary constants. The integrand of the D3 DBI action contains the factor
\beq
\mathcal{L}_{DBI,D3}= -e^{-\phi}\sqrt{ g+ F}\sim
\sqrt{16 r^2 +L^4(\cos^2\chi +\cos^2\zeta\sin^2\chi)\sin^4\zeta}.
\eeq
This encodes all the $\zeta$ dependent information about the D3-brane and is solved for $\zeta=0$. The action of the D3 brane is then given by
\begin{align}
S_{D3}&= -T_3\int d^4\sigma_1 e^{-\phi}\sqrt{-\det(g+ 2\pi F-B_2)}+ T_3 \int \mathcal{L}_{WZ D3}\\[3mm]
&=\frac{k_5 L^2}{8\pi}\int dt\left( \rho^3- \sqrt{1+\rho^2}\sqrt{\frac{16M^2 \pi^2}{L^4}+ \rho^4}\right),
\end{align}
and it  is once more minimized by
\beq
\r=0,~~~ \r=\frac{4 M\pi}{L^2},
\eeq
where the second of these is the fuzzy sphere solution. For either solution
\beq
E= \frac{k_5 M}{2}.
\eeq
Performing the analogous calculation for the D5 brane on  $\sigma_2^i=(t,S^2,r, \tilde{S}^2)$ leads to
\beq
E= \frac{k_3 M}{2}
\eeq
which once more shows that $k$ is replaced in the NATD theory by $k_3$ and $k_5$, with $k_5$ associated to D3 and $k_3$ associated to D5 branes. It is likely that these objects are related through duality to dual giant graviton configurations in M-theory, as in \cite{Nishioka:2008ib}.

We would now like to consider fuzzy sphere like solutions that propagate in the internal space. These are genuine Type II dual giant gravitons, that should be dual to singlet operators of the bifundamental matter fields \cite{Nishioka:2008ib}. Once more following \cite{Nishioka:2008ib} in ABJM (see also  \cite{Hamilton:2009iv})
     we  consider a D2 brane with embedding $\sigma^i=(t,S^2)$, however this time we fix the internal coordinates as
\beq
\phi_1 =\phi_1(t),~~~\zeta,\theta_1,\theta_2,\phi_2,\psi,\rho= {\rm constant}.
\eeq
The action of the D2-brane is given by
\beq
S_{D2}= \frac{kL^2}{8\pi} \int dt \,\Bigl(\r^3-\rho^2\sqrt{1+\r^2 -\cos^2\zeta (\sin^2\theta_1 + \sin^2\zeta \cos^2\theta_1)\dot{\phi}_1^2}\Bigr)\, .
\eeq 
The equations of motion for $\zeta,\theta_1,\theta_2$ are solved for two cases: $\zeta=\frac{\pi}{4}$, $\theta_1=0$ and $\zeta=0$, $\theta_1=\frac{\pi}{2}$, which reduce in the end to the same solution. 
A Lagrangian describing the motion of the D2-brane may then be written as
\beq
L_{D2} = \frac{kL^2}{8\pi}\int dt\, \rho^2 \Bigl(\r-\sqrt{1+\r^2-\dot{\phi}_1^2}\Bigr)\, .
\eeq
The Hamiltonian reads
\beq
H=P_{\phi_1} \dot{\phi}_1-L = \frac{kL^2}{8\pi}\left(\sqrt{1+\r^2}\sqrt{\frac{P^2_{\phi_1}}{a^2 }+\r^4}-\rho^3\right)\, ,
\eeq
and it is minimized by 
\beq
\r=0,~~~ \r= \frac{8\pi P_{\phi_1}}{kL^2}.
\eeq
The first of these is a graviton and the second a giant graviton solution. The energy of both  configurations is
\beq
E= P_{\phi_1}.
\eeq

We would now like to find an equivalent fuzzy sphere like solution propagating  in the T-dual internal space. As before, we naively expect the D2 brane to be mapped to a D3 extended in $\sigma^i_1=(t,S^2,r)$ and a D5 extended in $\sigma^i_2=(t,S^2,r,\tilde{S}^2)$.   
Since objects charged under the $SU(2)$ used to construct the NATD will be projected out by the transformation we consider a D3 brane  in the T-dual solution with $\sigma_1$ embedding and
\beq
\phi_1 =\phi_1(t),~~~\zeta,\theta_1,\chi,\xi,\r={\rm constant}.
\eeq
The induced metric and pullback of $B_2$ are then given by
\begin{align}
ds_{str,D3}^2&= \frac{L^2}{4}\bigg[-(1+\r^2-\Xi_1\dot{\phi}_1^2)dt^2+\r^2ds^2_{S^2}+\Xi_2 d r^2\bigg]\\[3mm]
B_{2,D3}&= \Xi_3 \dot{\phi}_1 \, dt\wedge dr
\end{align}
where
\begin{align}
 \Xi_1&=\cos^2{\zeta}\sin^2{\theta_1}+\frac{L^2 r^2 \sin^2{\chi} \cos^2{\theta_1}\sin^4{\zeta}\cos^2{\zeta}}{4 \det M}\nn\\[3mm]
 \Xi_2&=\frac{16r^2+ L^4(\cos^2\chi+\cos^2\zeta\sin^2\chi)}{4 \det M L^2}\\[3mm]
\Xi_3&=\frac{L^2 \cos^2\zeta\sin^2\zeta \cos\theta_1\cos\chi(16 r^2+L^2\sin^4\zeta)}{64 \det M}\nn
\end{align}
Solving the DBI action of the D3-brane for $\zeta,\chi,\theta_1$ we find $\zeta=0$, $\theta_1=0, \pi/2$.
Setting $\zeta=0$ in the DBI + WZ action leads to 
\beq
S_{D3}= \frac{k_5L^2}{8\pi} \int dt \r^2\left(\r-\sqrt{1+\r^2-\sin\theta_1^2 \dot{\phi}_1^2}\right)\, .
\eeq
Clearly only the $\theta_1= \frac{\pi}{2}$ case propagates and so we find, repeating the earlier calculation, that we have giant graviton solutions with 
\beq
\rho= \frac{8\pi P_{\phi_1}}{k_5 L^2}, ~~~E_{D3}= P_{\phi_1}.
\eeq
An analogous calculation for a D5 with embedding $\sigma_2$ and
\beq
\phi_1 =\phi_1(t),~~~\zeta,\theta_1,\r={\rm  constant}
\eeq
leads after $n$ large gauge transformations to 
\beq
\rho= \frac{8\pi P_{\phi_1}}{k_3 L^2},~~~ E_{D5}= P_{\phi_1}.
\eeq

Dual giant gravitons are thus also doubled in the NATD theory.
These configurations should  correspond in the CFT to singlet operators of the bifundamental matter fields associated to the two  gauge groups that arise in the NATD for each $U(N)$.
This is in line with our suggestion in the previous section for two sets of bifundamental matter fields for each pair of dual gauge groups with opposite levels.

\section{Central charge and entanglement entropy}

Three field theoretic objects which have a nice holographic description are the central charge \cite{Girardello:1998pd,Freedman:1999gp}, entanglement entropy of the strip \cite{Ryu:2006bv,Ryu:2006ef} and the free energy on the sphere \cite{Emparan:1999pm,Drukker:2010nc}. These all depend on the T-duality invariant quantity $e^{-2\phi}\sqrt{|det{g}|}$ (as well as on parts of the metric which are spectators under the duality). 
Thus the qualitative behaviour of these objects must be preserved by the $SU(2)$ isometry T-duality \cite{Itsios:2013wd}. We will see that they differ in fact by an unphysical constant common factor proportional to the volume of the $M_1\times {\tilde S}^2$ dual space.

Let us first look at the central charge and entanglement entropy. The specifics of the computations are given quite generally in \cite{Klebanov:2007ws}. We quote only the salient features and refer the interested reader to that reference for further details.

For a metric of the form,
\beq
ds^2_{str}= \alpha \beta d\rho^2+ \alpha dx_{1,d}^2+g_{ij}dy^idy^j,
\eeq
one can define the quantities
\beq\label{eq: quantitites}
\begin{array}{l l}
& V_{int}= \int d^{8-d}y \,e^{-2\phi}\sqrt{\det (g_{ij})} = \frac{1}{6}e^{-2\phi}L^6\pi^3\\ [2 mm]
& H= V_{int}^2\, \alpha^d = \frac{1}{576} e^{-4\phi}L^{16} \pi^6 r^4\\  [2 mm]
&\alpha= \frac{L^2 r^2}{4},~~~\beta =\frac{1}{r^4},~~~ d=2,~~~\kappa^d=H,
\end{array}
\eeq
where we quote their values  with respect to eqs (\ref{eq: metricbefore1}) and (\ref{eq: metricbefore2}).
This is a slight generalization of \cite{Klebanov:2007ws}, specifically in $V_{int}$, that allows the dilaton to depend on $y^i$, as it does for the T-dual solution. 

After the $SU(2)$ transformation reading off from the metric (\ref{metric1}), (\ref{dualmetric}), and dilaton (\ref{dilaton})  we define the new quantities
\beq
\begin{array}{l l}
&\tilde{V}_{int} =\int d^6\, \tilde{y}\, e^{-\tilde{\Phi}} \sqrt{det(\tilde{g}_{ij})} = \frac{1}{72} e^{-2\phi} L^6 (1+3n +3 n^2)\pi^5\\ [3 mm]
&\tilde{H}= \tilde{V}_{int}^2\alpha^2 
\end{array}
\eeq
where $\tilde{y}^i = (\zeta,\theta_1,\varphi_1,r,\chi,\xi)$ and we have integrated over $r\in [n\pi,(n+1)\pi)$. All other terms in eq (\ref{eq: quantitites}) remain unchanged.

The central charge of ABJM may then be read off as
\beq
c\sim \beta^{d/2} \kappa^{3d/2}(\kappa')^{-d}=\frac{1}{96} k^2 L^6 \pi^3 =\frac{4}{3} \sqrt{2} \sqrt{k} N^{3/2}\pi^6,
\eeq
while that of the T-dual is given by
\beq
\tilde{c} \sim \frac{1}{1152} k^2 L^6(1+3n +3n^2)\pi^5\, .
\eeq

\noindent Therefore, 
\begin{equation}
\label{relc}
\tilde{c}=\frac{\pi^2}{12}(1+3n +3n^2) \, c\, .
\end{equation}
We thus find that the central charges of the field theories dual to the solution for $r\in [n\pi, (n+1)\pi)$ differ from the central charge of  ABJM by a factor that depends on the range of $r$. In terms of the D3 and D5 Page charges we have
\beq
\label{centralc}
\tilde{c} \sim \frac{\sqrt{2}}{9}  \left(12 + \frac{N_5^2}{\left(N_3+\frac{N_5}{2}\right)^2}\right)\sqrt{k_5} \left(N_3 +\frac{N_5}{2}\right)^{3/2}\pi^6.
\eeq
Here we can see that the overall $\sqrt{k}N^{3/2}$ behaviour is essentially unchanged. These are common themes we will see with the other quantities we consider in this section.
For the 2-node field theory dual to the solution for $r\in [0,\pi)$ (\ref{centralc}) particularises to
\begin{equation}
{\tilde c}\sim \frac43 \sqrt{\frac23}\sqrt{k_3}\,N_5^{3/2}\pi^6=\frac{\pi^2}{12}\, c\, ,
\end{equation}
in agreement with (\ref{relc}) for $n=0$.

Let us now consider the entanglement entropy of an infinite strip of width $L_{EE}$, given, as in \cite{Klebanov:2007ws}, by solving the following parametric equations
\begin{align}
&L_{EE}(\rho_*)= 2 \sqrt{ H_*} \int_{\rho_*}^{\infty}\frac{ d\rho \sqrt{\beta}}{\sqrt{H-H_*}},\nn \\[3mm]
&S_{EE}(\rho_*)=S_{conn}-S_{diss}\sim \int_{\rho_*} d\rho\sqrt{\beta H}\big(\frac{\sqrt{ H}}{\sqrt{H-H_*}}-1\big)-\int_{0}^{\rho_*} d\rho\sqrt{\beta H},
\end{align}
where $H_*= H_{\rho=\rho_*}$. For ABJM this gives
\beq
L_{EE}(\rho_*) = \frac{2\sqrt{\pi}\Gamma[\frac{3}{4}]}{ \Gamma[\frac{1}{4}] \rho_*},~~~ S_{EE}(\rho_*) = - \frac{k^2 L^6\pi^{7/2} \Gamma[\frac{3}{4}]}{24 \Gamma[\frac{1}{4}]}\rho_*.
\eeq
Plugging the first of these into the second we arrive at \cite{Nishioka:2008gz}
\beq
S_{EE} = - \frac{k^2 L^6 \pi^4 \Gamma[\frac{3}{4}]^2}{12\Gamma[\frac{1}{4}]^2 L_{EE}}= - \frac{32\sqrt{2}\Gamma[\frac{3}{4}]^2 \sqrt{k} N^{3/2}\pi^7}{3 \Gamma[\frac{1}{4}]^2 L_{EE}}\, .
\eeq
Notice that, unlike the entanglement entropy of $AdS_5\times S_5$ (which goes like $S_{EE}\sim \frac{1}{L^2}$), we have $S_{EE}\sim \frac{1}{L}$. This is due to the lower dimensionality of the QFT.

The entanglement entropy in the T-dual solution can be shown to be
\beq
\tilde{S}_{EE} = - \frac{8\sqrt{2}\Gamma[\frac{3}{4}]^2\left(12+ \frac{N_5^2}{\left(N_3+\frac{N_5}{2}\right)^2}\right) \sqrt{k_5} \left(N_3 +\frac{N_5}{2}\right)^{3/2}\pi^7}{9\Gamma[\frac{1}{4}]^2 L_{EE}},
\eeq
where we once more see the behaviour is essentially unchanged between ABJM and its dual. Writing this expression in terms of the original $N,k$ parameters we find the same relation between the entanglement entropy of the field theory dual to the solution for $r\in [n\pi,(n+1)\pi)$ and the one for ABJM that we found for the central charges
\begin{equation}
\label{entropies}
\tilde{S}_{EE}=\frac{\pi^2}{12}(1+3n +3n^2) \, S_{EE}\, .
\end{equation}
For the field theory dual to the solution for $r\in [0,\pi)$ we find
\begin{equation}
\tilde{S}_{EE}=-\frac{32\sqrt{2} \Gamma[\frac{3}{4}]^2 \sqrt{k_3}N_5^{3/2}\pi^7}{3\sqrt{3}\Gamma[\frac{1}{4}]^2 L_{EE}}=\frac{\pi^2}{12}S_{EE}\, ,
\end{equation}
in agreement with (\ref{entropies}). 

Finally let us turn our attention to the free energy on $S^3$.  This may be computed using the results of \cite{Emparan:1999pm,Drukker:2010nc,Herzog:2010hf} which show that the free energy in $AdS_4$ is given by
\beq
F(S^3)= \frac{\pi L^2}{2 G_N}.
\eeq
$G_N$ is the 4d Newton constant which is given in terms of the 10d Newton constant and the volume of the internal space by
\beq
\frac{1}{G_N}= V_{int} \frac{1}{G_{10}} \, .
\eeq
In our conventions $G_{10} = (2 \pi)^6$, which leads to the free energy in ABJM \cite{Drukker:2010nc}
\beq
F(S^3)= \frac{\sqrt{2} \sqrt{k} N^{3/2} \pi}{3}\, .
\eeq
\noindent After the duality transformation the free energy is given by
\beq
\tilde{F}(S^3)=\frac{ \sqrt{2}\left(12+ \frac{N_5^2}{\left(N_3+\frac{N_5}{2}\right)^2}\right) \sqrt{k_5}\left( N_3+\frac{N_5}{2}\right)^{3/2} \p}{36}.
\eeq
Again the behaviour is unchanged between ABJM and its dual. In terms of the original $N,k$ parameters we find once more
\begin{equation}
\label{free}
\tilde{F}(S^3)=\frac{\pi^2}{12}(1+3n +3n^2) \, F(S^3)\, .
\end{equation}
Finally, for the field theory dual to the solution for $r\in [0,\pi)$ we find
\begin{equation}
\tilde{F}(S^3)=\frac{\sqrt{2}\sqrt{k_3}N_5^{3/2}\pi}{3^{3/2}}=\frac{\pi^2}{12}F(S^3)\, ,
\end{equation}
in agreement with (\ref{free}).  

All three quantities computed in this section thus differ from those in ABJM by an $n$-dependent common factor proportional to the volume of the $M_1\times {\tilde S}^2$ dual space 
\begin{equation}
{\rm Vol} (M_1\times {\tilde S}^2)=\int_{n\pi}^{(n+1)\pi} r^2 \sin{\chi}drd\chi d\xi=\frac43 (3n^2+3n+1)\pi^4\, .
\end{equation}
As we have already encountered, this factor originates because the volumes of the $S^3$ and $M_1 \times {\tilde S}^2$ manifolds spanned by the original and dual variables  violate the condition for the tensions $T_{p-3}=(2\pi)^3 T_p$, that is implicit in the calculations. Thus one should seek invariance for the quantities per unit volume, which are indeed the same in the original and dual theories.

\section{Conclusions}

An important drawback of non-Abelian T-duality is that it cannot be used to extract global information about the newly generated background. This is of foremost importance in backgrounds with CFT duals, where the existence of non-compact directions can pose serious problems to the CFT. Indeed, even if the group used to construct the NATD is compact, some of the newly generated variables, which now live in the Lie algebra, are not.

In most of the recent literature on non-Abelian T-duality this transformation has been used to generate new $AdS$ backgrounds from known ones containing an $S^3$, possibly squashed, in the internal space. Using as non-Abelian T-duality group a freely acting $SU(2)$, the $S^3$ gets  replaced  by  
$\mathbb{R}^3$, and a non-compact radial direction is generated in the dual geometry.

We have seen that for the NATD background considered in this paper, invariance under large gauge transformations  implies the existence of a flow in the dual CFT which allows one to interpret the running of $r$ very similarly to the cascade in \cite{Klebanov:2000hb}, with the important difference that the flow parameter is now the internal direction and not the energy scale, and thus relates different superconformal field theories. It is possible that  a similar interpretation is at work in other NATD backgrounds with CFT duals. 
 
We have argued that the end of the flow could be a $U(N_5)_{k_3}\times U(N_5)_{-k_3}$ quiver, which would be  the field theory dual to the solution for $r\in [0,\pi)$.  Based on the RR fields and branes that exist in the NATD background we can in fact speculate about a concrete brane realization in Type IIB.  Indeed, the way the RR fields transform
under NATD suggests that the original ABJM brane configuration \cite{Aharony:2008ug}
\begin{equation}
 \begin{array}{cc|lclclclclclclclc}
 \label{ABJM}
{\rm KK}:\ \ \ \    \   &   \times & \times &    \times &  \times &   \times  & \times &  z  & - &  - &   - \\
N\, {\rm D2}:\ \ \ \   \  &  \times & \times &  \times & -  & - & -  & -  & - &  -  & -   \\
({\rm KK},k\, {\rm D6}): \ \  &    \times & \times & \times & \cos{\theta} & \cos{\theta}& \cos{\theta} & \times & \sin{\theta} &\sin{\theta}& \sin{\theta}  
                         \end{array} 
\end{equation}
where $\theta$ is the angle between the KK-monopole and the $({\rm KK}, k\,D6)$-brane in the 3-7, 4-8 and 5-9 planes \footnote{Our notation for the $({\rm KK},k\, {\rm D6})$-brane tries to reflect explicitly that it is extended on the $x_1$, $x_2$ and $x_i \cos{\theta}+x_{i+4}\sin{\theta}$, with $i=1,2,3$, directions.},
is mapped into the following two configurations:
\begin{equation}
 \begin{array}{cc|lclclclclclclclc}
 \label{system1}
{\rm NS}5:\ \ \ \    \   &   \times & \times &    \times &  \times &   \times  & \times &  -  & - &  - &   - \\
N_3\, {\rm D3}:\ \ \ \   \ &   \times & \times &  \times & -  & - & -  & \times  & - &  -  & -   \\
({\rm NS5},k_5\,  {\rm D5}): \ \  &    \times & \times & \times & \cos{\alpha} & \cos{\alpha}& \cos{\theta} & - & \sin{\alpha} &\sin{\alpha}& \sin{\theta}  
                         \end{array} 
\end{equation}

\begin{equation}
 \begin{array}{cc|lclclclclclclclc}
 \label{system2}
5_2^2:\ \ \ \    \  &    \times & \times &    \times &  \times &   \times  & \times &  -  & z_1 &  z_2 &   - \\
N_5\, D5:\ \ \ \   \  &  \times & \times &  \times & -  & - & -  & \times  & \times &  \times  & -   \\
(5_2^2,k_3\,  {\rm D3}): \ \  &    \times & \times & \times & - & - &  \cos{\theta}  & - & - & - & \sin{\theta}  
                         \end{array} 
\end{equation}
when the duality is performed along the 6,7,8 directions\footnote{Here we have assumed that the 5-dimensional objects that appear after the NATD transformation of the KK-monopole coincide with the 5-branes that would be obtained after one or three Abelian T-duality transformations, namely a NS5-brane and a, so called, $5^2_2$ brane, respectively. This last object was identified in \cite{Hull:1997kb, Obers:1998fb}
as an exotic brane with two  isometric transverse directions (here we have used the notation of \cite{deBoer:2012ma}). It would be interesting to obtain the explicit transformation of the corresponding potentials under NATD to support this assumption.}. Note that the presence of the two   isometric directions $z_1$ and $z_2$ in the $5^2_2$-brane in the  (\ref{system2}) brane system reduces the dimensionality of the intersecting branes by 2, which agrees with the analysis of the quantized charges and the supersymmetries in the dual background. Moreover, consistency with $\mathcal{N}=2$ supersymmetry implies that the relative orientations in the 3-7 and 4-8 planes of the NS5 and $({\rm NS5}, k_5\, D5)$-brane in the  (\ref{system1}) brane system  should be different from that in the 4-9 plane after the duality. Our conjecture for the dual field theories in section 4.3 suggests then that 
 the field theory that describes the solution for $r\in[0,\pi)$ could be realized in the  D5-branes system described by (\ref{system2}),
given that at the end of the cascade the field theory seems to be a $U(N_5)_{k_3}\times U(N_5)_{-k_3}$ quiver, with both brane systems (\ref{system1}) and (\ref{system2})  playing a role in the realization of  the field theories that better describe the solution when $r\ge \pi$. 

The vanishing of the Kosmann derivative along the $SU(2)$ isometry on which we have T-dualized
suggests  that the new 3d CFT preserves $\mathcal{N}=2$ supersymmetry.
Although we have not provided a proof that one implies the other there is mounting evidence that this is indeed the case \cite{Itsios:2013wd,Barranco:2013fza,Macpherson:2013zba,Gaillard:2013vsa,Caceres:2014uoa}, and a proof for the situation for $SU(2)\subset SO(4)$\cite{Itsios:2012dc}. We have left a detailed proof for a general freely acting $SU(2)$ isometry for future work \cite{ENY}. Supersymmetry in Type II supergravity solutions with metric of the form $AdS_4\times M_6$, like those we consider here, has, on the other hand, a well understood geometric interpretation in terms of G-structures \cite{Grana:2005sn}. Specifically it is sufficient that there exists a closed NS-NS 3-form, non trivial RR sector and a $SU(3)\times SU(3)$-structure on the $M_6$ for some SUSY to be preserved \cite{Grana:2006kf}. The specific amount is determined by the number of such structures that may be defined for the same physical fields, with $SO(N)$ of them implying $\mathcal{N}=N $ in 2+1 dimensions. Although $\mathbb{CP}^3$ preserves an $SO(6)$ worth of $SU(3)$-structures, only a $U(1)$ of them are manifest in the Hopf fibration \cite{Gaiotto:2009yz}. As we expect $\mathcal{N}=2$ SUSY in the type IIB solution we should be able to define another $SO(2)$ worth of $SU(3)\times SU(3)$-structures, however it is possible that these would not all be manifest. The way T-duality works on the spinors by rotating the MW Killing spinors implies the new structure will be an $SU(2)$-structure. It would be interesting to check this explicitly and determine whether a full $SO(2)$ worth is manifest. It seems likely that the $SU(3)$-structures of the Hopf fibration that are manifest will get mapped to orthogonal $SU(2)$-structures under the pure spinor map proposed in \cite{Barranco:2013fza}, it would be gratifying to see this confirmed. 

As in the Type IIA dual of ABJM, we seem to have two $U(1)$ gauge fields for each of the gauge groups with positive levels in the dual CFT. For each gauge group, one $U(1)$ is associated to the RR symmetry of the 2-form (4-form) potential integrated on $M_1$ ($M_1 \times {\tilde S}^2$), and the other to the RR-symmetry of the 4-form (6-form) potential integrated over $M_1\times S^2_1$  ($M_1 \times S^2_1 \times {\tilde S}^2$). It would be interesting to see if one combination of these gauge fields is Higssed as in \cite{Aharony:2009fc}, such that in the end only one $U(1)$ remains for each of the dual gauge groups. One way to see this would be to look for configurations in the NATD theory with energy proportional to $k_3 N_5$ or $k_5 N_3$. It would be interesting to see as well if the uplift of our solution to 11 dimensions fits in the general classification of $\mathcal{N}=2$ $AdS_4$ solutions in 
\cite{Gabella:2012rc} \footnote{We would like to thank Dario Martelli for discussions on this point.}.

Finally, we would like to mention that we have ignored in this paper the subtle issue of the Freed-Witten anomaly  \cite{Freed:1999vc}, that affects some of the brane configurations of the original background \cite{Aharony:2009fc}. In particular, the di-baryons are realized in  $AdS_4\times \mathbb{CP}^3$ as D4-branes wrapped on a $\mathbb{CP}^2$. Given that the $\mathbb{CP}^2$ is not a spin manifold 
these branes have to carry a half-integer worldvolume field flux  in order to cancel the Freed-Witten anomaly. This flux should be compensated by a flat half-integer $B_2$ such that the field strength that couples in the brane worldvolume,  ${\cal F}=F-\frac{1}{2\pi}B_2$, is not modified, and the D4-brane does not lose its di-baryon interpretation. We have seen that the NATD theory shares many of the properties of the original ABJM theory. Therefore, one could expect some of the brane configurations that we have discussed in this paper to be also affected by the Freed-Witten anomaly. Our lack of knowledge of the global properties of the dual background does not allow us however to determine whether the sub-manifolds on which the dual branes are wrapped are spin or not. Still, even if a flat half-integer  $B_2$ field could modify the tadpole charges of some of the brane configurations that we have discussed, we would expect cancellations with higher curvature terms, as in the original ABJM theory  \cite{Aharony:2009fc}.

\subsection*{Acknowledgements}

We would like to thank Dario Martelli, Carlos Nu\~nez, Alfonso Ramallo,  Diego Rodr\'{\i}guez-G\'omez, Kostas Sfetsos and especially Eoin O Colgain for very useful discussions. This work has been partially supported by the 
COST Action MP1210 ``The String Theory Universe" with a Short Term Scientific Mission to Oviedo U.
Y.L. is partially supported by the Spanish Ministry of Science and Education grant FPA2012-35043-C02-02. N.T.M is supported by an STFC studentship. He is grateful for the warm hospitality extended by the hep-th group at Oviedo U., where this work was started.

\appendix

\section{Some further details of non-Abelian T-duality}

In this Appendix we include for completeness the expressions of the NATD RR field strengths derived in 
\cite{Itsios:2012zv} as well as the expressions for the dual RR field strengths ${\hat F}_7$ and ${\hat F}_9$.

Writing the RR field strengths of the original background as
\begin{equation}
F_p=G_p^{(0)}+G_{p-1}^a \wedge e^a+\frac12 G_{p-2}^{ab}\wedge e^a \wedge e^b+G_{p-3}^{(3)}\wedge e^1\wedge e^2\wedge e^3\, ,
\end{equation}
the ones in the NATD background\footnote{We use tildes to denote them throughout this section.}
\begin{equation}
\label{RRdual}
{\tilde F}_p={\tilde G}_p^{(0)}+{\tilde G}_{p-1}^a \wedge {\tilde e}^a+\frac12 {\tilde G}_{p-2}^{ab}\wedge {\tilde e}^a \wedge {\tilde e}^b+{\tilde G}_{p-3}^{(3)}\wedge {\tilde e}^1\wedge {\tilde e}^2\wedge {\tilde e}^3\, ,
\end{equation}
can be derived from the transformation rules
\begin{eqnarray}
&&{\tilde G}_{p}^{(0)}=\Bigl( -A_0G_p^{(3)}+A_a G_p^a \Bigr)\nonumber\\
&&{\tilde G}_{p-1}^a=\Bigl( -\frac{A_0}{2} \epsilon^{abc}G_{p-1}^{bc}+A_b G_{p-1}^{ab}+A_a G_{p-1}^{(0)}\Bigr) \nonumber\\
&&{\tilde G}_{p-2}^{ab}=\Bigl[ \epsilon^{abc}\Bigl( A_c G_{p-2}^{(3)}+A_0 G_{p-2}^c \Bigr)-(A_a G_{p-2}^b-A_b G_{p-2}^a)\Bigr] \nonumber\\
&&{\tilde G}_{p-3}^{(3)}=\Bigl(\frac{A_a}{2}\epsilon^{abc}G_{p-3}^{bc}+A_0 G_{p-3}^{(0)}\Bigr)\, .
\end{eqnarray}
Here $\phi$  and ${\tilde \phi}$ refer to the dilaton in the original and NATD theory, $a,b,c$ run over the directions of the squashed $S^3$ on which the dualization is performed, $e^a$ are the frames on these directions and the coefficients $A_0$ and $A_a$ are given by
\begin{equation}
A_0=e^{B_1+B_2+B_3}\, , \qquad A_a = v_a e^{B_a}\, ,
\end{equation}
where $A_a$ and $B_a$ are functions of the spectator coordinates, $v_i$ are the Lagrange multipliers living in the Lie algebra of $SU(2)$, and we have assumed an original geometry of the form
\beq
ds^2_{str} = g_{\mu\nu}dx^{\mu}dx^{\nu}+ e^{2B_a}(\omega^a +A^a)^2,
\eeq
where $\omega^a$ are left invariant 1-forms and $A^a = A^a_{~\mu}dx^{\mu}$ are fibration terms.

We include as well for reference the higher fluxes
\begin{align}
\tilde{ \hat{F}}_7 = d{\rm Vol}(AdS_4)\wedge \bigg[&\frac{1}{8}k L^6 \sin\zeta\cos^2\zeta {\rm Vol}(S^2_1)\wedge\left(\sin\zeta(\cos\chi dr + r\sin\chi d\chi)-2r \cos\zeta \cos\chi d\zeta\right)\nn \\
&-\frac{1}{4}k L^6 \cos\zeta \sin^4\zeta \cos\theta_1\left(\cos\chi d\zeta\wedge d\phi_1\wedge dr- r \sin\chi d\zeta\wedge d\phi_1\wedge d\chi\right) \nn\\
&- 6 kL^2 r^2 dr\wedge {\rm Vol}(\tilde{S}^2)\bigg]\\[3mm]
\tilde{\hat{F}}_9=-\frac{1}{4} k L^6\sin\zeta\cos\zeta^3&d{\rm Vol}(AdS_4)\wedge  {\rm Vol}(S^2_1)\wedge\bigg[r^2\sin\zeta \sin\chi d\zeta\wedge {\rm Vol}(\tilde{S}^2)-r \sin^2\chi d\zeta\wedge dr\wedge d\xi \bigg]\nn.
\end{align}

\section{Killing spinors of ABJM}\label{sec: AB}

In this Appendix we derive the Killing spinors on $\mathbb{CP}^3$. The work is not new, see for instance \cite{Hohenegger:2009as}, however to our knowledge this is the first time they are presented for the Hopf fibration of $\mathbb{CP}^3$ in the frame 
\begin{align}\label{eq:frame}
&e^{x^{\mu}}=\frac{L}{2} dx^{\mu},~~~ e^{r}=\frac{L}{2r}dr,~~~ e^1= L d\zeta,~~~ e^2= \frac{L}{2}\cos\zeta d\theta_1,~~~e^3 = \frac{L}{2}\cos\zeta\sin\theta_1 d\phi_1,\nn\\
& e^4 = \frac{L}{2}\sin\zeta\,\omega_1,~~~e^5 = \frac{L}{2}\sin\zeta\,\omega_1,~~~e^6 = \frac{L}{2}\sin\zeta\cos\zeta(\omega_3+\cos\theta_1 d\phi_1).
\end{align}
With respect to this frame the RR sector may be simply expressed as
\beq
F_2 = \frac{2k}{L}J,~~~J= -2(e^{16}+e^{23}+e^{45}),~~~ F_4 =\frac{6k}{L}e^{x^0x^1x^2r}
\eeq
and the spin connection is
\begin{align}\label{eq:spinconnection}
&\omega^{x^{\mu}r}= \frac{2}{L} e^{x^{\mu}},~~~\omega^{12}=-\omega^{36}=\frac{\tan\zeta}{L}e^{2},~~~\omega^{13}=\omega^{26}=\frac{\tan\zeta}{L}e^{3}\nn\\
& \omega^{14}=-\omega^{56}=-\frac{\cot\zeta}{L} e^4,~~~\omega^{15}=\omega^{46}=-\frac{\cos\zeta}{L}e^5,~~~\omega^{16}-\frac{2\cot 2\zeta}{L}e^6,\\
&\omega^{23}=\frac{1}{L} (-2\cot\theta_1\sec\zeta e^3+\tan\zeta e^6),~~~\omega^{45}=\frac{1}{L}(-2\cot\theta_1\sec\zeta e^3+(\cot\zeta+2\tan\zeta )e^6)\nn,
\end{align}
with all other components zero. One can show that  the spin connection given by (\ref{eq:spinconnection}) satisfies $de^A+\omega^A_{~B}\wedge e^{B}=0$.
 We will use the same gamma matrices as \cite{Bakhmatov:2010fp}, where the 10d matrices are built out of Kronecker products of 4 and 6 dimensional ones. The 4-dimensional matrices are given by
\beq
\begin{array}{ccccc}
\alpha^0 &=&\sigma^3&\otimes& i \s^2\\
\alpha^1 &=&\sigma^3&\otimes & \s^1\\
\alpha^2 &=&\sigma^3&\otimes & \s^3\\
\alpha^3 &=&\sigma^1&\otimes &1 
\end{array}
\eeq
where $\sigma^i$ are the Pauli matrices. The 6 dimensional matrices are
\beq
\begin{array}{ccccccc}
\gamma^1 &=&1&\otimes& \s^2&\otimes &\s^1\\
\gamma^2 &=&1&\otimes& \s^2&\otimes &\s^3\\
\gamma^3 &=&\s^1&\otimes& 1&\otimes& \s^2\\
\gamma^4 &=&\s^3&\otimes& 1&\otimes &\s^2\\
\gamma^5 &=&\s^2&\otimes& \s^1&\otimes& 1\\
\gamma^6 &=&\s^2&\otimes &\s^3&\otimes& 1.
\end{array}
\eeq
The real 10 dimensional gamma matrices are then
\beq
\Gamma^{\mu}= i \alpha^{\mu}\otimes 1,~~~~\Gamma^{i+3} = i\alpha^{(4)}\otimes \gamma^i
\eeq
where $\mu=0,1,2,3$ and $i=1,...6$. The 10 dimensional chirality matrix is given by
\beq
\Gamma^{(10)}=- \a^{(4)}\otimes \gamma^{(7)}
\eeq
where $\a^{(4)}=\a^0\a^1\a^2\a^3$ and $\gamma^{(7)}= - \gamma^1\gamma^2\gamma^3\gamma^4\gamma^5\gamma^6$.

The Killing spinor of $AdS_4\times \mathbb{CP}^3$ must satisfy the following conditions
\begin{align}
\delta\lambda &= \frac{e^{\phi}}{8} \bigg(\frac{3}{2!}F_{ab}\Gamma^{ab}\Gamma^{(10)}- \frac{1}{4!} F_{abcd}\Gamma^{abcd}\bigg)\epsilon=0\\
\delta_{M}\Psi &= \nabla_M\epsilon - \frac{e^{\phi}}{8}\bigg(\frac{1}{2!}F_{ab}\Gamma^{ab}\Gamma^{(10)}+\frac{1}{4!} F_{abcd}\Gamma^{abcd}\bigg)\Gamma_{M}\epsilon=0,
\end{align}
where $D_{M}\epsilon =\partial_{M}+\frac{1}{4}\omega_{M,ab}\Gamma^{ab}\epsilon$. These can be solved separately for the $AdS_4$ and $\mathbb{CP}^3$ parts of the space by introducing a 10 dimensional spinor that is a product of 4 and 6 dimensional pieces $\epsilon= \chi\otimes \eta$. As shown for instance  in \cite{Bakhmatov:2010fp,Nilsson:1984bj} the requirement that the dilatino variation vanishes is dependent on the 6d spinor only:
\beq
(Q+2)\eta=0\, ,
\eeq
where we define $Q= J_{ab}\gamma^{ab}\gamma^{(7)}$ as in \cite{Nilsson:1984bj}. It is simple to show that in our chosen frame $Q= diag(-2,-2,-2,6,6,-2,-2,-2)$ and so clearly the most general Killing spinor must be of the form
\beq
\eta= (\eta_1,\eta_2,\eta_3,0,0,\eta_4,\eta_5,\eta_6)^{T}, 
\eeq
where we must fix $\eta_i$ with the gravitino variation. The requirement that these vanish can also be split in 4+6 parts with the conditions on $\chi$ giving a standard differential condition for a spinor in $AdS_4$ and hence preservation of 4 supercharges. The condition on $\eta$ can be cast in the form
\beq
D_a\eta+\frac{i}{2L} \bigg(\gamma_a-J_{ab}\gamma^b\gamma^{(7)}\bigg)\eta=0.
\eeq
Specifically this gives the following set of coupled first order differential equations
\begin{align}\label{eq CP3spinor}
&\bigg(\partial_{\zeta}+\frac{i}{2}(\gamma^1+ \gamma^{12345})\bigg)\eta=0\\
&\bigg(4\sec\zeta\partial_{\theta_1}+i(\gamma^2-\gamma^{12456})+\tan\zeta(\gamma^{12}-\gamma^{36})\bigg)\eta=0\nn\\
&\bigg(2\csc\theta_1\sec\zeta\big(\cos\theta_1(2\partial_{\psi}+\gamma^{23}+\gamma^{45})-2\partial_{\phi_1}\big)-\tan\zeta(\gamma^{13}+\gamma^{26})-i(\gamma^3-\gamma^{13456})\bigg)\eta=0\nn\\
&\bigg(4\csc\theta_1\big(\sin\psi\partial_{\theta_2}+\cos\psi\csc\theta_2(-\partial_{\phi_2}+\cos\theta_2\partial_{\psi})\big)+\cot\zeta(\gamma^{14}-\gamma^{56})\big)-i(\gamma^4-\gamma^{12346})\bigg)\eta=0\nn\\
&\bigg(4\csc\zeta\big(\cos\psi\partial_{\theta_2}+\sin\psi(\csc\theta_2\partial_{\phi_2}-\cot\theta_2\partial_{\psi})\big)-\cot\zeta(\gamma^{15}+\gamma^{46})+i(\gamma^5-\gamma^{12356})\bigg)\eta=0\nn\\
&\bigg(8 \partial_{\psi}+\gamma^{23}+3\gamma^{45}-\cos2\zeta(2\gamma^{16}+\gamma^{23}+\gamma^{45})+i \sin2\zeta(\gamma^6+\gamma^{23456})\bigg)\eta=0\nn.
\end{align}
Solving these explicitly in Mathematica leads to
\begin{align}
\label{CP3spinors}
\eta_1&=\frac{1}{2} \bigg[2 \sin \left(\frac{\theta _1}{2}\right) \bigg(\cos \left(\frac{\theta _2}{2}\right) \left(c_4 \cos \left(\frac{1}{2} \left(\psi +\phi _1+\phi
   _2\right)\right)-c_3 \sin \left(\frac{1}{2} \left(\psi +\phi _1+\phi _2\right)\right)\right)-\nn\\
	&~~~\sin \left(\frac{\theta _2}{2}\right) \left(c_1 \sin \left(\frac{1}{2}
   \left(\psi +\phi _1-\phi _2\right)\right)+c_2 \cos \left(\frac{1}{2} \left(\psi +\phi _1-\phi _2\right)\right)\right)\bigg)+\nn\\
	&~~~2 \cos \left(\frac{\theta _1}{2}\right)
   \bigg(\cos \left(\frac{\theta _2}{2}\right) \left(c_2 \sin \left(\frac{1}{2} \left(\psi -\phi _1+\phi _2\right)\right)+c_1 \cos \left(\frac{1}{2} \left(\psi -\phi _1+\phi
   _2\right)\right)\right)-\nn\\
	&~~~\sin \left(\frac{\theta _2}{2}\right) \left(c_4 \sin \left(\frac{1}{2} \left(-\psi +\phi _1+\phi _2\right)\right)+c_3 \cos \left(\frac{1}{2}
   \left(-\psi +\phi _1+\phi _2\right)\right)\right)\bigg)\bigg]\nn\\[2mm]
	\eta_2&=\sin (\zeta ) \bigg[\cos \left(\frac{\theta _1}{2}\right) \bigg(\cos \left(\frac{\theta _2}{2}\right) \left(c_4 \cos \left(\frac{1}{2} \left(\psi +\phi _1+\phi
   _2\right)\right)-c_3 \sin \left(\frac{1}{2} \left(\psi +\phi _1+\phi _2\right)\right)\right)-\nn\\
	&~~~\sin \left(\frac{\theta _2}{2}\right) \left(c_1 \sin \left(\frac{1}{2}
   \left(\psi +\phi _1-\phi _2\right)\right)+c_2 \cos \left(\frac{1}{2} \left(\psi +\phi _1-\phi _2\right)\right)\right)\bigg)+\nn\\
	&~~~\sin \left(\frac{\theta _1}{2}\right)
   \bigg(\sin \left(\frac{\theta _2}{2}\right) \left(c_4 \sin \left(\frac{1}{2} \left(-\psi +\phi _1+\phi _2\right)\right)+c_3 \cos \left(\frac{1}{2} \left(-\psi +\phi _1+\phi
   _2\right)\right)\right)-\nn\\
&	~~~\cos \left(\frac{\theta _2}{2}\right) \left(c_2 \sin \left(\frac{1}{2} \left(\psi -\phi _1+\phi _2\right)\right)+c_1 \cos \left(\frac{1}{2}
   \left(\psi -\phi _1+\phi _2\right)\right)\right)\bigg)\bigg]+c_5 \cos (\zeta )\nn\\[2mm]
	\eta_3&=\cos (\zeta ) \bigg[\cos \left(\frac{\theta _1}{2}\right) \bigg(\sin \left(\frac{\theta _2}{2}\right) \left(c_1 \sin \left(\frac{1}{2} \left(\psi +\phi _1-\phi
   _2\right)\right)+c_2 \cos \left(\frac{1}{2} \left(\psi +\phi _1-\phi _2\right)\right)\right)+\nn\\
	&~~~\cos \left(\frac{\theta _2}{2}\right) \left(c_3 \sin \left(\frac{1}{2}
   \left(\psi +\phi _1+\phi _2\right)\right)-c_4 \cos \left(\frac{1}{2} \left(\psi +\phi _1+\phi _2\right)\right)\right)\bigg)+\nn\\
	&~~~\sin \left(\frac{\theta _1}{2}\right)
   \bigg(\cos \left(\frac{\theta _2}{2}\right) \left(c_2 \sin \left(\frac{1}{2} \left(\psi -\phi _1+\phi _2\right)\right)+c_1 \cos \left(\frac{1}{2} \left(\psi -\phi _1+\phi
   _2\right)\right)\right)-\nn\\
	&~~~\sin \left(\frac{\theta _2}{2}\right) \left(c_4 \sin \left(\frac{1}{2} \left(-\psi +\phi _1+\phi _2\right)\right)+c_3 \cos \left(\frac{1}{2}
   \left(-\psi +\phi _1+\phi _2\right)\right)\right)\bigg)\bigg]+c_5 \sin (\zeta )\nn\\[2mm]
	\eta_4&=\cos (\zeta ) \bigg[\sin \left(\frac{\theta _1}{2}\right) \bigg(\cos \left(\frac{\theta _2}{2}\right) \left(c_2 \cos \left(\frac{1}{2} \left(\psi -\phi _1+\phi
   _2\right)\right)-c_1 \sin \left(\frac{1}{2} \left(\psi -\phi _1+\phi _2\right)\right)\right)+\nn\\
	&~~~\sin \left(\frac{\theta _2}{2}\right) \left(c_4 \cos \left(\frac{1}{2}
   \left(-\psi +\phi _1+\phi _2\right)\right)-c_3 \sin \left(\frac{1}{2} \left(-\psi +\phi _1+\phi _2\right)\right)\right)\bigg)+\nn\\
	&~~~\cos \left(\frac{\theta _1}{2}\right)
   \bigg(\sin \left(\frac{\theta _2}{2}\right) \left(c_1 \cos \left(\frac{1}{2} \left(\psi +\phi _1-\phi _2\right)\right)-c_2 \sin \left(\frac{1}{2} \left(\psi +\phi _1-\phi
   _2\right)\right)\right)+\nn\\
	&~~~\cos \left(\frac{\theta _2}{2}\right) \left(c_4 \sin \left(\frac{1}{2} \left(\psi +\phi _1+\phi _2\right)\right)+c_3 \cos \left(\frac{1}{2}
   \left(\psi +\phi _1+\phi _2\right)\right)\right)\bigg)\bigg]-c_6 \sin (\zeta )\nn\\[2mm]
	\eta_5&=\sin (\zeta ) \bigg[\sin \left(\frac{\theta _1}{2}\right) \bigg(\cos \left(\frac{\theta _2}{2}\right) \left(c_2 \cos \left(\frac{1}{2} \left(\psi -\phi _1+\phi
   _2\right)\right)-c_1 \sin \left(\frac{1}{2} \left(\psi -\phi _1+\phi _2\right)\right)\right)+\nn\\
	&~~~\sin \left(\frac{\theta _2}{2}\right) \left(c_4 \cos \left(\frac{1}{2}
   \left(-\psi +\phi _1+\phi _2\right)\right)-c_3 \sin \left(\frac{1}{2} \left(-\psi +\phi _1+\phi _2\right)\right)\right)\bigg)+\nn\\
	&~~~\cos \left(\frac{\theta _1}{2}\right)
   \bigg(\sin \left(\frac{\theta _2}{2}\right) \left(c_1 \cos \left(\frac{1}{2} \left(\psi +\phi _1-\phi _2\right)\right)-c_2 \sin \left(\frac{1}{2} \left(\psi +\phi _1-\phi
   _2\right)\right)\right)+\nn\\
	&~~~\cos \left(\frac{\theta _2}{2}\right) \left(c_4 \sin \left(\frac{1}{2} \left(\psi +\phi _1+\phi _2\right)\right)+c_3 \cos \left(\frac{1}{2}
   \left(\psi +\phi _1+\phi _2\right)\right)\right)\bigg)\bigg]+c_6 \cos (\zeta )\nn\\[2mm]
	\eta_6&=\frac{1}{2} \bigg[2 \cos \left(\frac{\theta _1}{2}\right) \bigg(\cos \left(\frac{\theta _2}{2}\right) \left(c_2 \cos \left(\frac{1}{2} \left(\psi -\phi _1+\phi
   _2\right)\right)-c_1 \sin \left(\frac{1}{2} \left(\psi -\phi _1+\phi _2\right)\right)\right)+\nn\\
	&~~~\sin \left(\frac{\theta _2}{2}\right) \left(c_4 \cos \left(\frac{1}{2}
   \left(-\psi +\phi _1+\phi _2\right)\right)-c_3 \sin \left(\frac{1}{2} \left(-\psi +\phi _1+\phi _2\right)\right)\right)\bigg)-\nn\\
	&~~~2 \sin \left(\frac{\theta _1}{2}\right)
   \bigg(\sin \left(\frac{\theta _2}{2}\right) \left(c_1 \cos \left(\frac{1}{2} \left(\psi +\phi _1-\phi _2\right)\right)-c_2 \sin \left(\frac{1}{2} \left(\psi +\phi _1-\phi
   _2\right)\right)\right)+\nn\\
	&~~~\cos \left(\frac{\theta _2}{2}\right) \left(c_4 \sin \left(\frac{1}{2} \left(\psi +\phi _1+\phi _2\right)\right)+c_3 \cos \left(\frac{1}{2}
   \left(\psi +\phi _1+\phi _2\right)\right)\right)\bigg)\bigg],
\end{align}
where $c_i$ are the 6 arbitrary constants required for the solution to preserve $\mathcal{N}=6$ supersymmetry in 3 dimensions.
These are clearly extremely complicated expressions, however notice that all the dependence on $\theta_1,\phi_1,\theta_2,\phi_2,\psi$ lies within the square bracketed terms. These can all be set to zero by fixing $c_1=c_2=c_3=c_4=0$ leaving a simplified spinor which depends only on $\zeta$ and 2 arbitrary constants. It is this fact that will be central to our argument that some supersymmetry is preserved in the non-Abelian T-dual.

\end{document}